\documentclass[14pt]{article}

\usepackage{graphics}
\usepackage{graphicx}
\usepackage{amssymb}
\usepackage{amsmath}
\usepackage{array}
\usepackage{hhline}
\usepackage{multirow}
\usepackage{verbatim}
\usepackage{slashed}
\usepackage{multirow}

\usepackage{cite}

\newcommand{\redim}[2]{
	\addtolength{\voffset}{-#1}
	\addtolength{\textheight}{#1}
	\addtolength{\textheight}{#1}

	\addtolength{\hoffset}{-#2}
	\addtolength{\textwidth}{#2}
	\addtolength{\textwidth}{#2}
}
\newcommand{\B}{ {B}}
\newcommand{\T}{ {T_{5/3}}}
\newcommand{\MT}{ M_{T}}
\newcommand{\mt}{ {m_T} }
\newcommand{\mtt}{ {m_{T2}} }
\newcommand{\MB}{ M_{\B}}
\newcommand{\ET}{ {\slashed{E}_T}}
\newcommand{\PT}{ {p_T}}
\newcommand{\HT}{ {H_T}}
\newcommand{\kt}{ {k_T}}
\newcommand{\MGME}{\textsc{MadGraph/MadEvent}}
\newcommand{\Pythia}{\textsc{Pythia}}
\newcommand{\getjet}{\textsc{GetJet}}

\redim{2cm}{2cm}

\title{
\vspace{0.0 cm}
{\huge
A Strong Sector at the LHC:\\ \vspace*{0.3cm} 
Top Partners in Same--Sign Dileptons
}
\vspace{0.4cm}
\author{{\Large \text{Jan Mrazek}\footnote{jan.mrazek@epfl.ch},\, \text{Andrea Wulzer}\footnote{andrea.wulzer@epfl.ch}} \\ \\
{\emph{ITP, EPFL, CH-1015, Lausanne, Switherland}}}
}

\date{}
\begin{document}
\maketitle \thispagestyle{empty}

\begin{center}
{\large Abstract} \\

\vspace{1cm}

\parbox[c]{12cm}{
Heavy partners of the top quark are a common prediction of many models in which a new strongly--coupled sector is responsible for the breaking of the EW symmetry. In this paper we investigate their experimental signature at the LHC, focusing on the particularly clean channel of same--sign dileptons.

We show that, thank to a strong interaction with the top quark which allows them to be singly produced at a sizable rate, the top partners will be discovered at the LHC if their mass is below $1.5$~TeV, higher masses being possible in particularly favorable (but plausible) situations. Being the partners expected to be lighter in both the Higgsless and Composite--Higgs scenarios, the one of same--sign dileptons is found to be a very promising channel in which these models could be tested.

We also discuss several experimental signatures which would allow, after the discovery of the excess, to uniquely attribute it to the top partners production and to measure the relevant physical parameters, {\it{i.e.}} the top partners masses and couplings. We believe that our results constitute a valid starting point for a more detailed experimental study.
}

\end{center}
\newpage

\section{Introduction}

Models in which a new strongly coupled sector is responsible for the breaking of the electroweak symmetry (EWSB), solving the Hierarchy Problem, have received renewed attention in the last few years. Progresses came from warped compactifications \cite{Randall:1999ee} which allowed to reformulate old scenarios such as Technicolor \cite{Weinberg:1975gm} and Composite--Higgs \cite{Dugan:1984hq} in terms of calculable five--dimensional (5d) effective theories leading respectively to the Higgsless \cite{Csaki:2003dt-bulk, Cacciapaglia:2006gp, Chivukula:2006cg-bulk} and to the Minimal Composite Higgs models \cite{Contino:2003ve-bulk, Contino:2006qr}.

That any of these 5d models is exactly dual to some 4d strong dynamics is far from established, but it is known that striking similarities exist. Compatibly with our qualitative understanding of strongly--coupled dynamics any 5d model result can be consistently interpreted in 4d language. For phenomenological purposes, however, the 5d models could be considered per s{\'e} and their validity as effective field theories might well extend above the maximum energy reach of the LHC. Moreover, their UV completion could come from string theory rather than from a strong sector. If this is the case, the strongly--coupled language would just be a useful tool to give an alternative (and sometimes simpler) interpretation to the 5d--theory results.

The 5d models not only provided calculable realization of previous scenarios, they also suggested new model--building solutions. It is the case of the ``partial compositeness'' idea (originally proposed in \cite{Kaplan:1991dc}) which is automatically implemented in the 5d construction \cite{Contino:2004vy} and actually constitutes its key feature in terms of which much of the 5d physics can be captured by simple 4d models \cite{Contino:2006nn}. In partial compositeness the SM fermions $f$ (similarly to the SM vector bosons which mix with the currents) couple to the strong sector, and therefore acquire their mass after EWSB, by mixing linearly with some strong--sector operator ${\mathcal{O}}$, {\it{i.e.}} through terms like $f\,{\mathcal{O}}$ in the UV Lagrangian. In the IR, where a mass--gap is generated, a composite fermion with the quantum numbers of ${\mathcal{O}}$ exists and the UV term is converted into a mixing of the SM fermion with this composite state. After EWSB, the mass eigenstates are a light SM fermion and its heavy partner, with mass of the TeV order. Since small masses require small mixings, the light SM families are mostly elementary and  very weakly coupled to the strong sector. This realizes the so--called RS-GIM mechanism of flavor protection \cite{Gherghetta:2000qt} which almost, but not completely \cite{Csaki:2008zd}, solves the long--standing flavor problem of previous strong--sector EWSB models. Among the partners, the one of the top quark is special and is the subject of our study. To acquire its high mass, indeed, the top quark must have a sizable composite component and therefore sizable interactions with the strong sector and in particular with its partner. These interactions could make deviations from the SM of the top interactions detectable \cite{Kumar:2009vs-bulk} and they play a major role in our study of the top partners production at the LHC.

The existence of top partners is a generic and robust prediction of the 5d models and, more generally, of 5d--inspired partial compositeness scenarios \cite{Contino:2006nn}. Another signature are massive vector resonances in the adjoint of the color group; the latter are expected to be present if the strong sector carries color as it must in partial compositeness in order for the strong sector operators ${\mathcal{O}}$ to mix with the quarks. These resonances are the partners of the gluon in the language of \cite{Contino:2006nn} and arise as Kaluza--Klein (KK) gluons in the 5d models, their production and decay to tops should be observable at the LHC if their mass is below about $4$~TeV \cite{Agashe:2006hk-bulk}. The most direct and robust signature of strong--sector EWSB (independently on whether it realizes partial fermion compositeness or not) would however be the detection of color--singlet vectors with Electro--Weak (EW) quantum numbers, essentially analog to the QCD $\rho$ mesons. The latter are the partners (or the KK, in the 5d language) of the EW gauge bosons and play a crucial role in the $WW$ scattering unitarization even in the case of a composite Higgs at high enough energies. These states will be visible at the LHC only if they are lighter than about $2$ or $3$~TeV \cite{Birkedal:2004au}. The EW partners, however, directly contribute to the $S$ parameter at tree--level and in the absence of cancellations their mass should be higher than a few ($\gtrsim3$) TeV in order for the model to be compatible with precision EW measurements 
\cite{Carena:2007ua-bulk,Contino:2006qr}. In the higgsless model the EW partners are forced to be light, and therefore possibly visible at the LHC, since they are the sole responsible for $WW$ unitarization. This requires a fine--tuning in $S$ which renders the model less appealing. \footnote{It is actually the request of an enhanced calculability ({\it{i.e.}} of a large number of colors $N_c$) which, combined with $WW$ unitarization, implies low EW partners mass. In a QCD--like case ($N_c=3$) the partners should be as heavy as $2.5$~TeV and a moderate tuning would be sufficient.} In the composite Higgs case, on the contrary, the EW partners masses (or, which is the same, the strong--sector compositeness scale $\Lambda$) can be above $3$~TeV without violating perturbative unitarity due to the presence of an Higgs particle which can postpone unitarity violation to higher energies. This is achieved by a fine--tuning in $v/f$, where $v=246$~GeV is the Higgs VEV and $f$ is the decay constant of the Higgs, which arises as a Goldstone boson in these models. Making $f$ large, the EW partners become heavy and no extra fine--tuning is needed in $S$. The EW partners, however, will be invisible at the LHC and it becomes difficult, more in general, to probe the strongly--coupled nature of the Higgs sector\cite{Giudice:2007fh-bulk}.

In this paper we study the possibility of observing the top partners in the extremely clean channel of two same-sign hard and separated leptons, large total transverse energy $H_T$ and some missing energy $\ET$. We will show that discovery is possible, if the coupling to tops is large as expected, up to $1.5$~TeV top partner mass, but it becomes difficult if they are heavier. The channel we study is therefore relevant for the higgsless case, in which the strong scale $\Lambda$ is low and it is natural to have top partners below $1.5$~TeV, while it might appear marginal in models with $\Lambda$ of about $3$~TeV. This is not  the case, however, in the compelling composite Higgs scenario where, as discussed in detail in \cite{Contino:2006qr}, the top partners are always parametrically lighter than $\Lambda$ and lie in the $[.5,1.5]$~TeV mass range. This happens in all the allowed parameter space and independently on details of the model such as the 5d representations in which the SM fermions are embedded. An heuristic explanation of this is that the Higgs mass--term, which is finite and calculable in these models, is the result of a cancellation between the low--energy SM--like contribution and the high--energy contribution of the new states. The bigger SM contribution comes from the top loops and is canceled by the top partners loops (which play in this context essentially the same role as the stop loops in supersymmetry) while the SM gauge fields loops are canceled by the EW partners contribution. 
But the Higgs is light (below $190$~GeV) in these models, which is also helpful with EWPT, and 
requiring an upper bound on $m_h$ enforces an upper bound on the mass of the partners, which 
have to be light enough for the SM divergence cancellation to begin at small enough energies. The 
most stringent bound is on the top partners, since their role is to cancel the larger SM contribution, 
while the EW partners are allowed to be heavier. Along the lines above, the bound of $1.5$~TeV 
on the top partners mass can be derived, and a correlation of the top partners mass with the Higgs 
mass is established, we refer the reader to \cite{Contino:2006qr} for a complete 
discussion. Summarizing, top partners in the $[.5,1.5]$~TeV range are likely to be the best experimental signature of the composite Higgs scenario, all other new states being expected sensibly heavier.

The production of heavy colored fermions at the LHC has been extensively studied \cite{Atre:2008iu-bulk, Han:2003wu-bulk}
, but the analysis which is more closely related to ours is the one of ref.~\cite{Contino:2008hi}, which also considered top partners in same--sign dileptons. While ref.~\cite{Contino:2008hi} focused on pair production, we also consider the single--production mediated by the previously discussed interaction with the top quark. Being the latter the distinctive feature of the top partner, the inclusion of this production channel will help in distinguishing it from a generic colored heavy fermion and will permit, as we will discuss, a simple measurement of the coupling. Moreover, the single production greatly enhances the cross--section for high masses and makes discovery possible in the entire range of interest. At the technical level, our event--selection strategy differs from the one of \cite{Contino:2008hi}. We indeed find an $H_T$ cut to be extremely efficient in reducing the background. Our selection makes only use of this variable, of the leptons momenta and of some $\ET$. We include in our analysis the experimental effect of charge misidentification which is potentially important for same--sign dileptons due to the very large opposite--sign SM background. We make the conservative assumption of $1\%$ flat charge misidentification probability (typical lepton $\PT$ $100<\PT<500$ GeV) and with this assumption one main background is an opposite--sign process. Our results could therefore improve if charge misidentification is closer to $10^{-3}$ as expected. We also discuss how, after the discovery of the excess, the underlying exotic particle content could be identified and the masses and couplings of the top partners measured, we believe that our result are a valid starting point for a more detailed experimental analysis.

\section{The Model}

We describe the top partners using the language of partial compositeness, that provides a simple parametrization of 5d strong--sector EWSB models \cite{Contino:2006nn}. We therefore introduce heavy vector--like colored fermions $Q$ and $\widetilde{T}$ which are the partner of, respectively, the standard $q_L=(t_L,b_L)$ doublet and the $t_R$ singlet; we ignore light families and the $b_R$ partners since they will not play a role in what follows. The strong sector, which generates the partners, is assumed to respect a global $G_s=SU(2)_L\times SU(2)_R \times U(1)_X$ symmetry, where the SM $SU(2)_L$ is embedded in the first factor while for the Hypercharge we have $Y=T_{3}^R+X$. The first $SU(2)_L\times SU(2)_R\simeq SO(4)$ factor will be spontaneously broken down to the custodial $SO(3)_c$, either by the strong sector itself as in the Higgsless scenario or by the Higgs VEV in the composite Higgs, making in both cases purely strong--sector contributions to the T parameter vanish. The partners therefore live in representations of $G_s$, and since their role is to give a mass to the top quark  they must be chosen such that a $G_s$--invariant ``proto--Yukawa" term for them exists and at the same time they can mix with the SM fermions without breaking the SM group. This last requirement actually forces the strong sector to carry color as an additional global symmetry, the group is $SU(3)_c\times G_s$ and the partners are color triplets.

Both cases in which the strong sector delivers an Higgs field or not can be treated simultaneously if we write
\begin{equation}
H\,=\,\left[\begin{array}{cc}h_d^\dagger & h_u\\ -h_u^\dagger & h_d\end{array}\right]\,=\,{ \frac{v}{\sqrt{2}}}\,U\,
\simeq\,\left[\begin{array}{cc} \frac{1}{\sqrt{2}}\left(v-i\varphi_0\right) & \varphi_+\\
                                                     - \varphi_-     & \frac{1}{\sqrt{2}}\left(v+i\varphi_0\right)
                      \end{array}\right]\,,
                      \label{Ghiggs}
\end{equation}
where $H$ is the Higgs field, in the $(\bf{2},\bf{2})_0$ of $G_s$, $U$ is the Goldstone bosons unitary matrix parametrized by the neutral ($\varphi_0$) and charged ($\varphi_-={\varphi_+}^\dagger$) Goldstone fields and the last approximate equality is obtained by expanding $U$ at the first order in the Goldstones. In the case of a Composite Higgs, $v$ should be a dynamical degree of freedom whose fluctuations describe the physical Higgs boson, but since we are not interested in interactions of the partners with the physical Higgs we have set it to its VEV $v=246$~GeV. We will write the proto--Yukawa interactions in term of $H$, but after making use of eq.~(\ref{Ghiggs}) we will obtain the same Lagrangian we would have written in the higgsless case where only the Goldstones in $U$ (and not the entire $H$) are present. What we denote as proto--Yukawa term is actually, in the higgsless case, a non--$SO(4)$ invariant mass term properly ``dressed'' with Goldstones in order for the $SO(4)$ symmetry to be restored.

The concrete model we consider is the same as in \cite{Contino:2008hi}, which provides a parametrization of the composite Higgs model of \cite{Contino:2006qr} and of the higgsless model of \cite{Cacciapaglia:2006gp}. The top partner representations and the associated proto--Yukawa term are
\begin{equation}
Q=
({\bf{2}},{\bf{2}})_{2/3}=
\left[
\begin{array}{cc}
T & T_{5/3}\\
B & T_{2/3}
\end{array}
\right]\,,
\;\;\;\;\;\;
{\widetilde{T}}=({\bf{1}},{\bf{1}})_{2/3}\,,
\;\;\;\;\;\;
{\mathcal L}_Y=Y_t^* {\textrm{Tr}}\left[ \overline{Q} H\right] \widetilde{T}+\textrm{h.c.}\,,
\label{lag0}
\end{equation}
where the $(T,B)$ doublet has the same SM quantum numbers as $q_L=(t_L,b_L)$ and $\widetilde{T}$ the ones of $t_R$. The $Q$ and $\widetilde{T}$ multiplets have masses $M_{Q,\widetilde{T}}$ of the order (though a bit smaller in the composite Higgs case, as we have discussed) of the compositeness scale $\Lambda\sim$~TeV, and also mix with strength ${\Delta}_{Q,\widetilde{T}}$ to $q_L$ and $t_R$. 
Diagonalizing  the mixings one gets a mass term for the top. From eq.~(\ref{lag0}) we find
\begin{equation}
y_t\,=\,\frac{\sqrt{2}m_t}v\,=\,Y_t^*\,\sin{\varphi_q}\,\sin{\varphi_t}\,,
\label{mt}
\end{equation}
where $\varphi_{q,t}$ are the $q_L,t_R$ mixing angles.

The equation above immediately tells us that the proto--Yukawa coupling $Y_t^*$ cannot be very small, it has at least to exceed $y_t\simeq1$. It will actually be bigger in concrete models, because of the following. If $Y_t^*$ is generated by a strong dynamics (or by an extra--dimensional model) it can be estimated as
$$
Y_t^*\,=\,\frac{4\pi}{\sqrt{N}}\,,
$$
where $N$ is the number of colors of the strong sector. In the 5d language, $1/\sqrt{N}$ corresponds to the expansion parameter and indeed making $N$ big makes our IR description of the strong sector more weakly coupled. A strong bound on the number of colors $N$ comes from the S parameter which grows linearly with $N$ as a result of the fact that the vector resonances mass $m_\rho$ decreases (at fixed $4\pi f\sim m_\rho\sqrt{N}$) as $1/\sqrt{N}$. This requires $N\lesssim10$ \cite{Contino:2006qr} which in turn implies $Y_t^*\gtrsim4$. As an upper bound on $Y_t^*$, which could still ensure calculability in the 5d model, we could take for instance the $g_{\rho}$ coupling of the $\rho$ meson of QCD which corresponds to $N=3$ and is around $6$.

\begin{figure}[!tb]
	\begin{center}
	\begin{tabular}{ccc}
		$\B$	&& $\T$ \\&&\\
		\scalebox{0.39}{\includegraphics{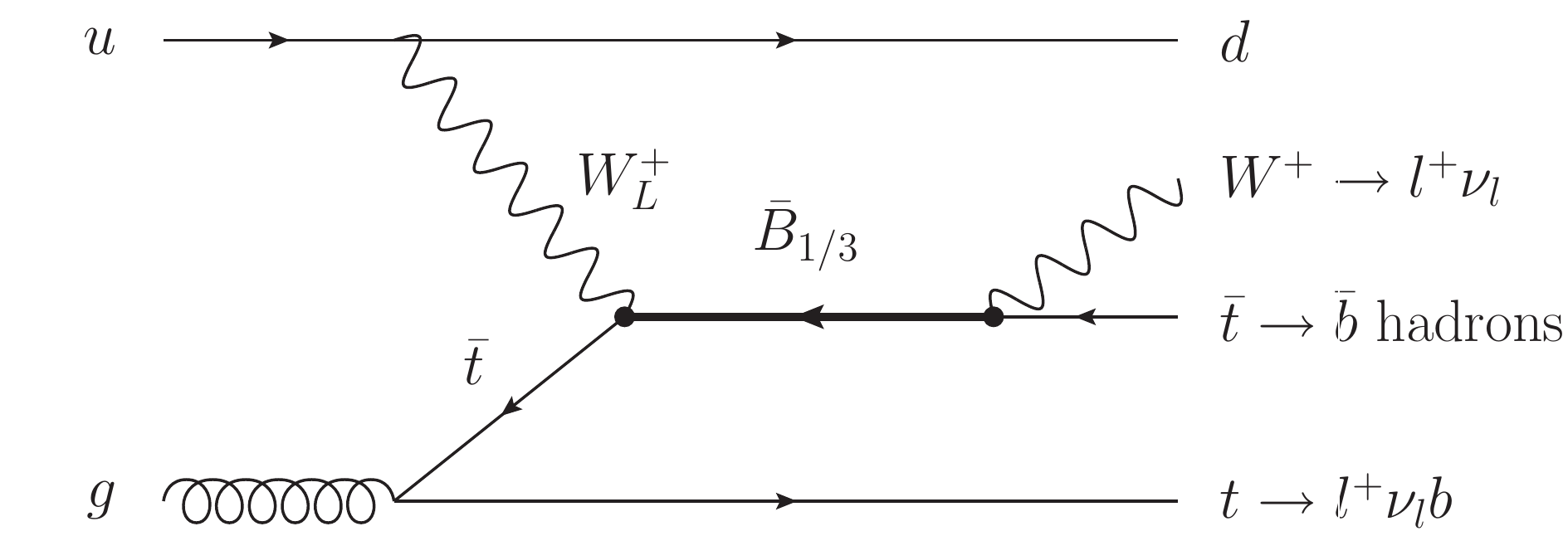}}
		&&
		\scalebox{0.39}{\includegraphics{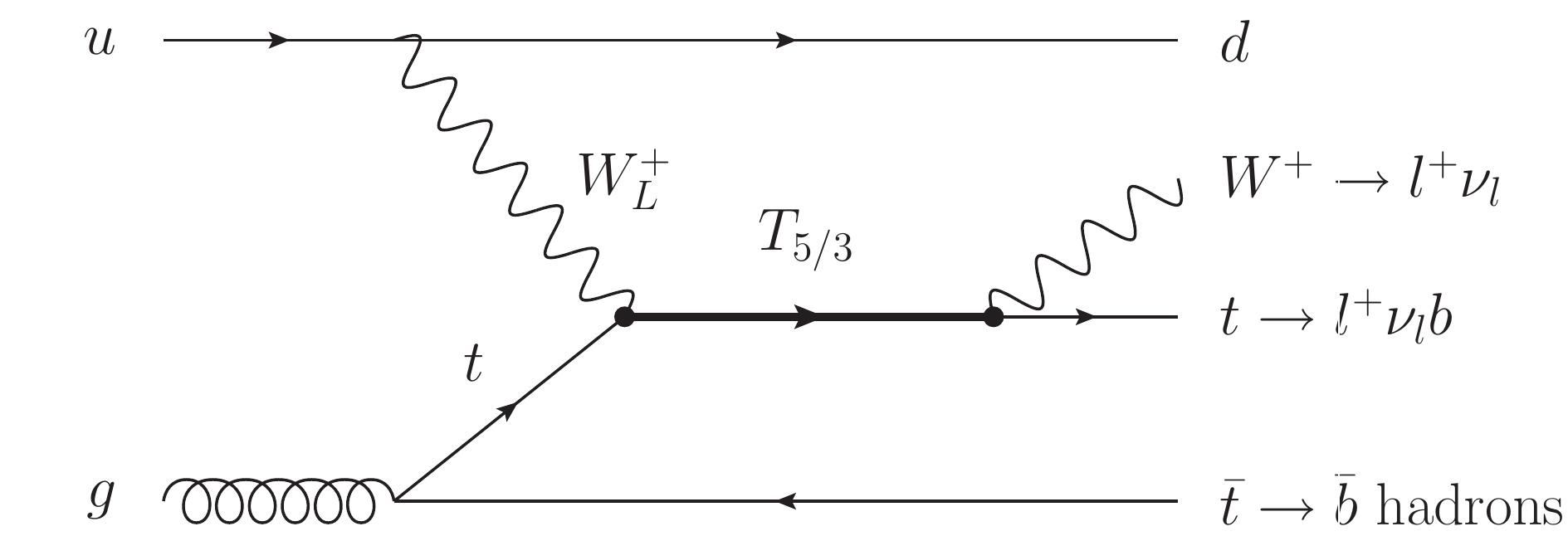}} \\ &&\\
		\scalebox{0.39}{\includegraphics{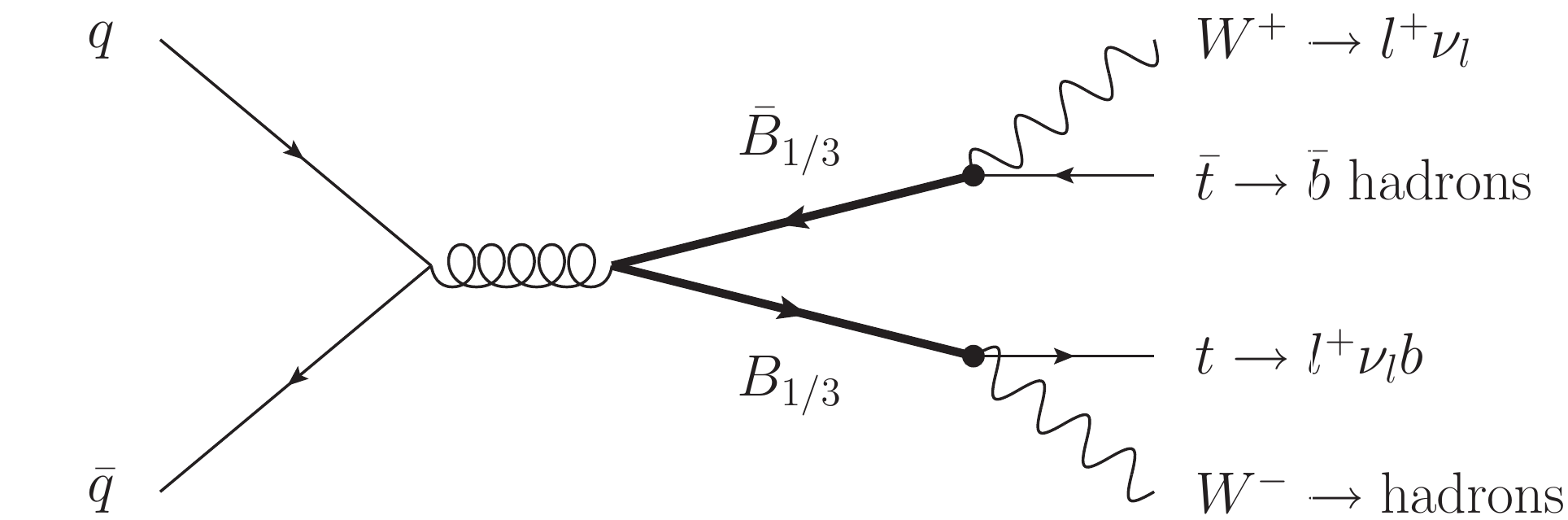}}
		&&
		\scalebox{0.39}{\includegraphics{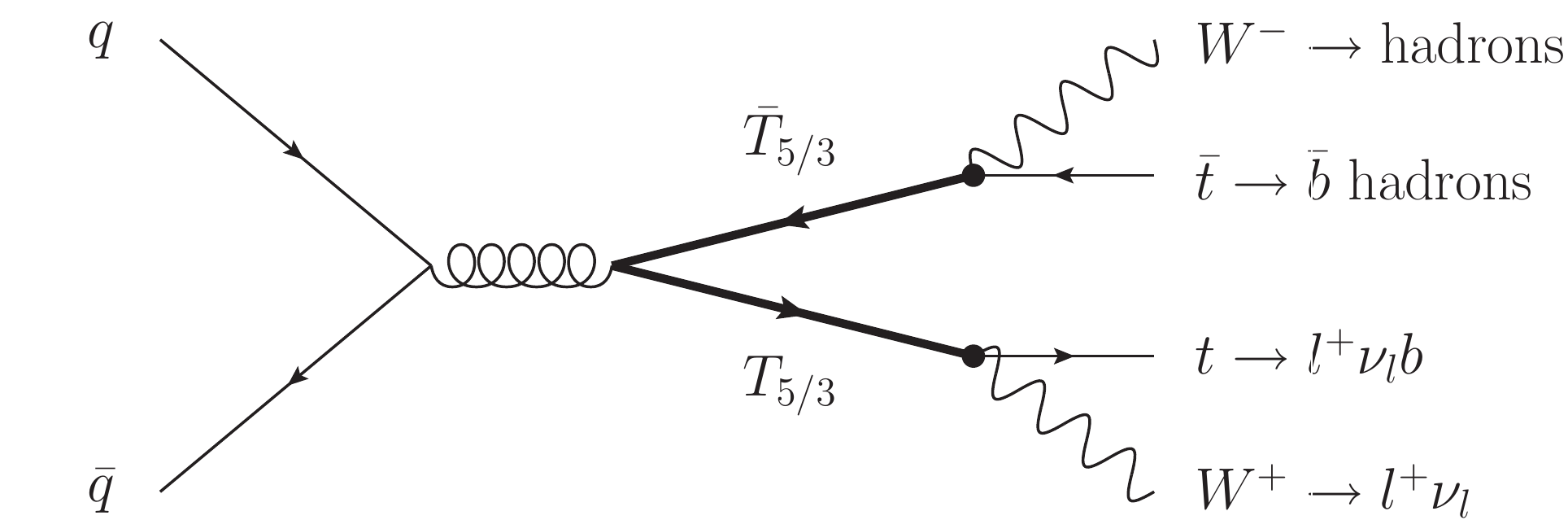}}
	\end{tabular}
	\end{center}
	\caption{Typical single and pair production diagrams for $\T$ and $\B$ for signals with two positively charged leptons.
		We notice that for $\T$ the leptons always comes from its decay, while for $\B$ they originate in two different legs.}
	\label{figProdDiagram}
\end{figure}

The lagrangian in eq.~(\ref{lag0}) also delivers top partners interactions with two SM particles, which will mediate top partner decay and single production. These are
\begin{eqnarray}
{\mathcal L}_Y=&& - Y_t^* \sin{\varphi_t}\cos{\varphi_q} \varphi_{+} \overline{t}_R B\,+\,
 Y_t^*  \sin{\varphi_t}\varphi_{-} \overline{t}_R T_{5/3}\,+\,
i\,Y_t^*  \sin{\varphi_t}\cos{\varphi_q}\frac{\varphi_0}{\sqrt{2}} \overline{t}_R T\,\nonumber\\
\ &&-i\, Y_t^* \sin{\varphi_t}\frac{\varphi_0}{\sqrt{2}}  \overline{t}_R T_{2/3}\,-\,
Y_t^* \sin{\varphi_q}\cos{\varphi_t}\left[
\varphi_{-}  \overline{b}_L +i \frac{\varphi_0}{\sqrt{2}}\overline{t}_L
\right]
\widetilde{T}\,+\,\textrm{h.c.}
\,,
\label{lagc}
\end{eqnarray}
and correspond, when going to the unitary gauge and making use of the Equivalence Theorem, to vertices with the longitudinal EW bosons. From the Lagrangian above it is easy to see that only the $B$ and the $T_{5/3}$ partners will be visible in the final state we want to study, which contains two hard and separated same--sign leptons; the pair and single production diagrams are shown in fig.~\ref{figProdDiagram}.

The couplings $\lambda_B=Y_t^* \sin{\varphi_t}\cos{\varphi_q}=y_t/\tan{\varphi_q}$ and $\lambda_T=Y_t^* \sin{\varphi_t}=y_t/\sin{\varphi_q}$ are potentially large since $Y_t^*$ is large, as we have discussed, and for sure $\lambda_T\geq y_t\simeq1$. 
But they will actually be bigger in realistic models where the amount of compositeness of $q_L$, $\sin{\varphi_q}$, cannot be too large. 
The $b_L$ couplings have indeed been measured with high precision and showed no deviations from the SM. 
Large $b_L$ compositeness would have already been discovered, for instance in deviations of the $Z b_L\overline{b}_L$ coupling from the SM prediction. Generically, corrections $\delta g_L/g_L\sim \sin{\varphi_q}^2\left( v/f \right)^2$ \cite{Contino:2006nn} are expected which would imply (for moderate tuning $v/f\;{ \slash \hspace{-10pt}\ll}\,1$) an upper 
bound on $\sin{\varphi_q}$. It is however possible to eliminate such contributions by imposing, as in the model 
of \cite{Contino:2006qr} (see also \cite{Carena:2006bn-bulk}), a ``Custodial Symmetry for $Z b_L\overline{b}_L$'' \cite{Agashe:2006at} 
which makes the correction reduce to $\delta g_L/g_L\sim \sin{\varphi_q}^2\left(m_Z/\Lambda\right)^2$. Still, having 
not too big $b_L$ compositeness is favored and further bounds are expected to come from flavor constraints in the 
$B$--meson sector. To be more quantitative we can assume that $\sin{\varphi_q}<\sin{\varphi_t}$, {\it{i.e.}} 
that $q_L$ is less composite than the $t_R$. 
This implies $\sin{\varphi_q}<\sqrt{(y_t/Y_t^*)}$ and therefore $\lambda_T>\sqrt{(y_t Y_t^*)}\gtrsim 2$ and $\lambda_B>\sqrt{(y_t Y_t^* - y_t^2)}\gtrsim \sqrt{3}$. 
We will therefore consider $\lambda_{T,B}$ couplings which exceed 2 and use the reference values of $2,3,4$; smaller values for both couplings are not possible under the mild assumption $\sin{\varphi_q}<\sin{\varphi_t}$.

Our analysis, though performed in the specific model we have described, has a wide range of applicability. The existence of the $B$ partner is, first of all, a very general feature of the partial compositeness scenario given that one partner with the SM quantum numbers of the $b_L$ must exist. Also, it interacts with the $t_R$ as in eq.~(\ref{lagc}) due to the $SU(2)_L$ invariance of the proto--Yukawa term. The $T_{5/3}$ could on the contrary not exist, this would be the case if for instance we had chosen representations $Q=({\bf{2}},{\bf{1}})_{1/6}$ and $\widetilde{T}=({\bf{1}},{\bf{2}})_{1/6}$ for the partners (which is however strongly disfavored by combined bounds from $\delta g_b/g_b$ and T), or in the model of \cite{Contino:2006nn}. To account for these situations we will also consider the possibility that only the $B$ partner is present. \footnote{In this case, our analysis perfectly applies to the model proposed in \cite{Contino:2006nn}, where the $t_R$ is entirely composite, $\sin{\varphi_t}=1$, and the coupling is large.} The existence of the $T_{5/3}$ is a consequence of the $Z b_L\overline{b}_L$--custodial symmetry, which requires that the $B$ partner has equal $T^3_L$ and $T^3_R$ quantum number. This, plus the $SO(4)$ invariance of the proto--Yukawa, implies that the $T_{5/3}$ must exist and couple as in eq.~(\ref{lagc}). Our analysis, as we have remarked, can also apply to Higgsless scenarios in both cases in which the custodian $T_{5/3}$ is present or not. The results could change quantitatively in other specific models because for instance other partners can be present and contribute to the same--sign dilepton signal, or other channels could open for the decay of the partners making the branching ratio to top, which is one in our model, decrease. This cannot however qualitatively invalidate our conclusions on the discovery, which are robust if the partners are not too heavy and their couplings, which determine the single production cross--section, are not too small. 

\section{Discovery Analysis}
\begin{figure}[!htb] 
	\begin{center}
		\scalebox{0.4}{\rotatebox{0}{\includegraphics{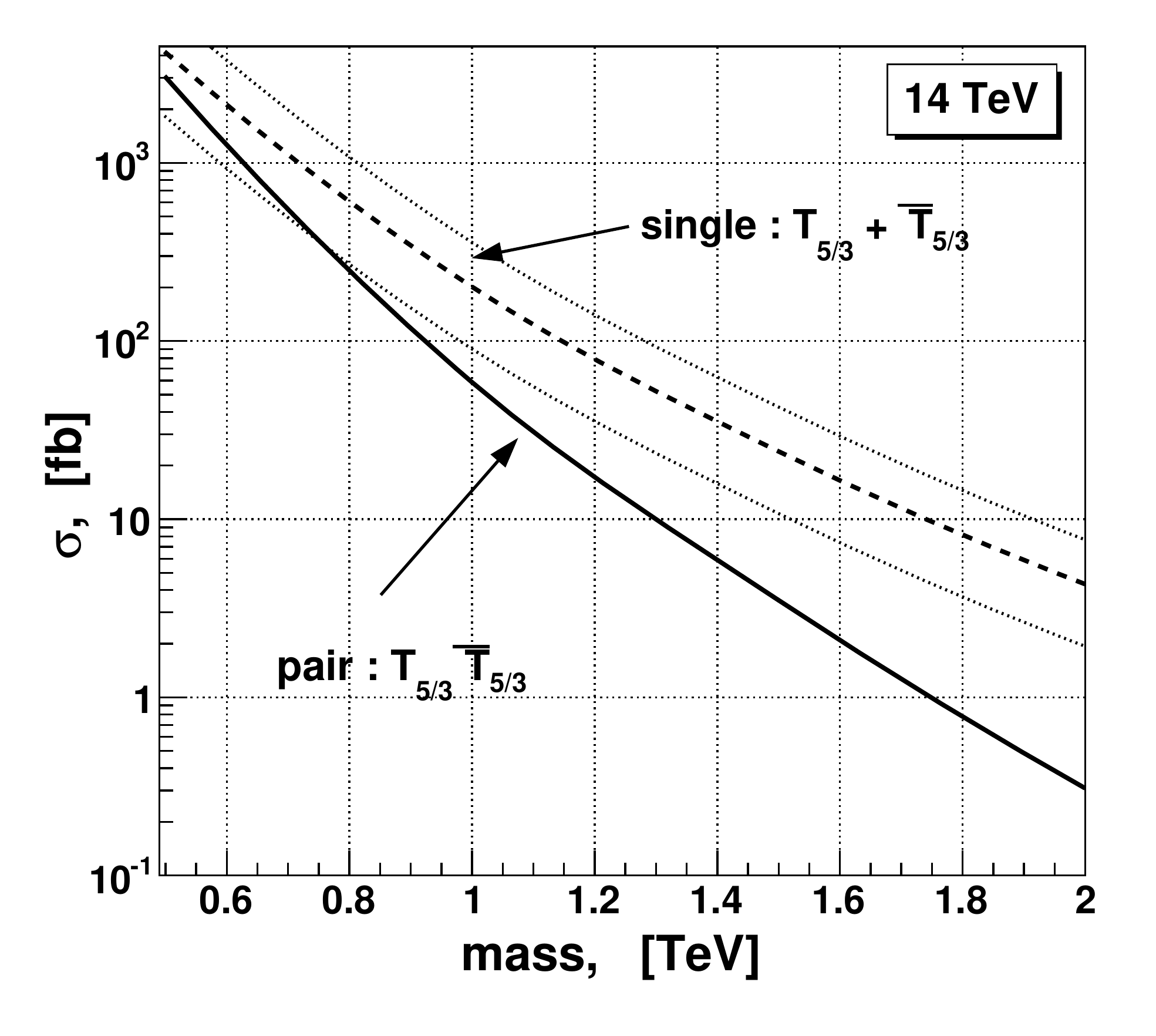}}}
	\end{center} 
	\caption{Cross sections, summed over charge, for pair (plain) and
		single (dashed) production of $\T$ (or $B$) as a function of its mass.  The
		dotted lines show the effect for the single production of varying
		$2<\lambda_{T,B}<4$.}
	\label{figXsec}
\end{figure}

The cross--sections of single and pair top partners production at the LHC are shown in fig.~\ref{figXsec}. After production, the partners decay to top quark and $W$ as depicted in fig.~\ref{figProdDiagram} with unit branching ratio, but reaching the dilepton final state will cost us an extra factor of $\approx \frac{2}{9}\cdot\frac{2}{9}\cdot\frac{6}{9}\approx 0.03$ (id.$\cdot \frac{6}{9}\approx0.02$) for single (pair) production.
Compared with pair, the single production cross--section is always sizable in
the mass range we are interested in and, since it decreases slower with the top
partner mass,  rapidly becomes dominant. This is somewhat surprising since the
single production diagram contains one weak interaction vertex and is also
suppressed, in comparison with pair production, by the three--body phase
space. Very similar situations, however, are encountered in the case of a fourth heavy
family production, studied in \cite{Willenbrock:1986cr}, and in the phenomenology 
of little Higgs models \cite{Han:2003wu-bulk}.

\begin{figure}
	\centering
	\begin{tabular}{cc}
		\rotatebox{90}{\scalebox{0.36}{\includegraphics{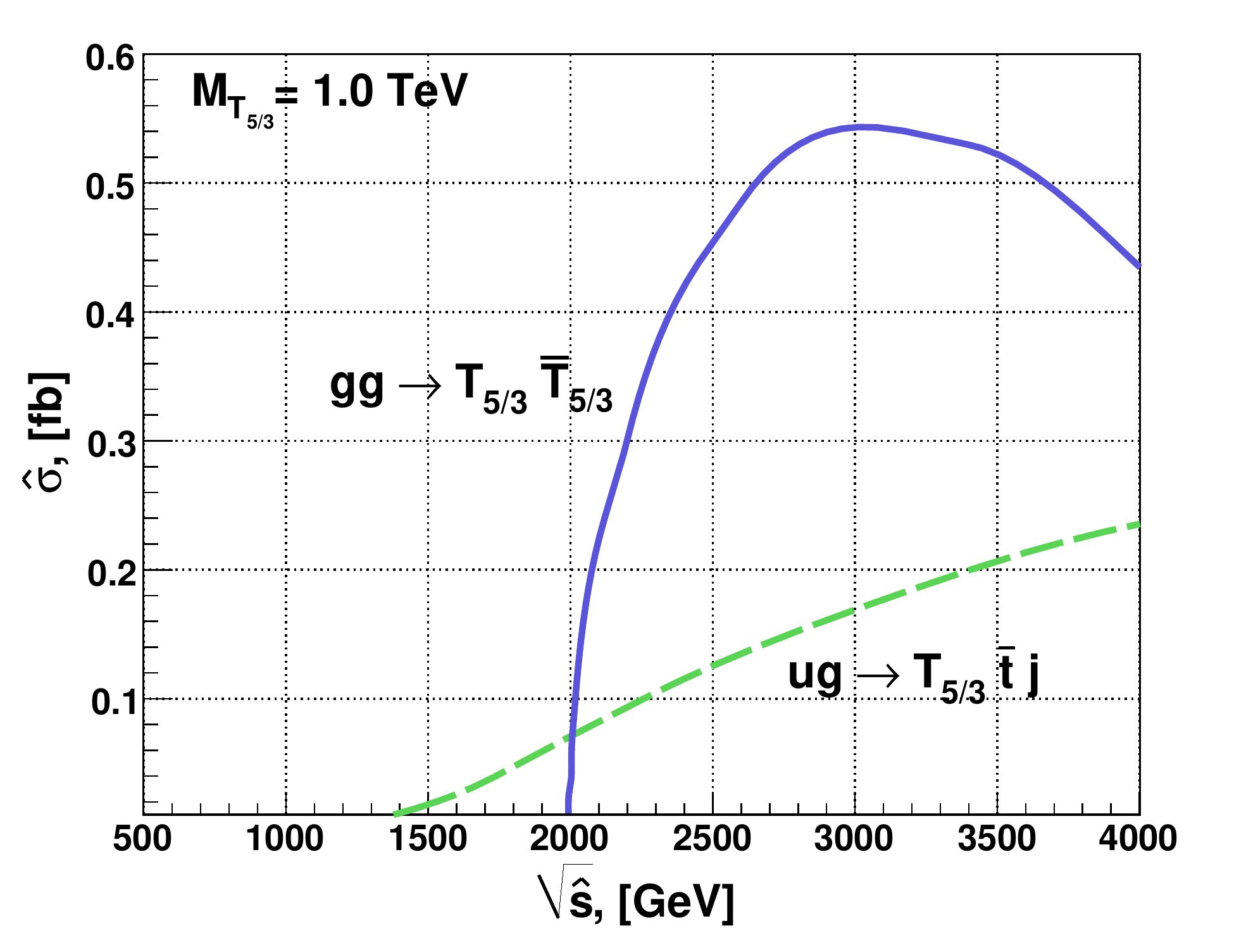}}} &
		\rotatebox{90}{\scalebox{0.36}{\includegraphics{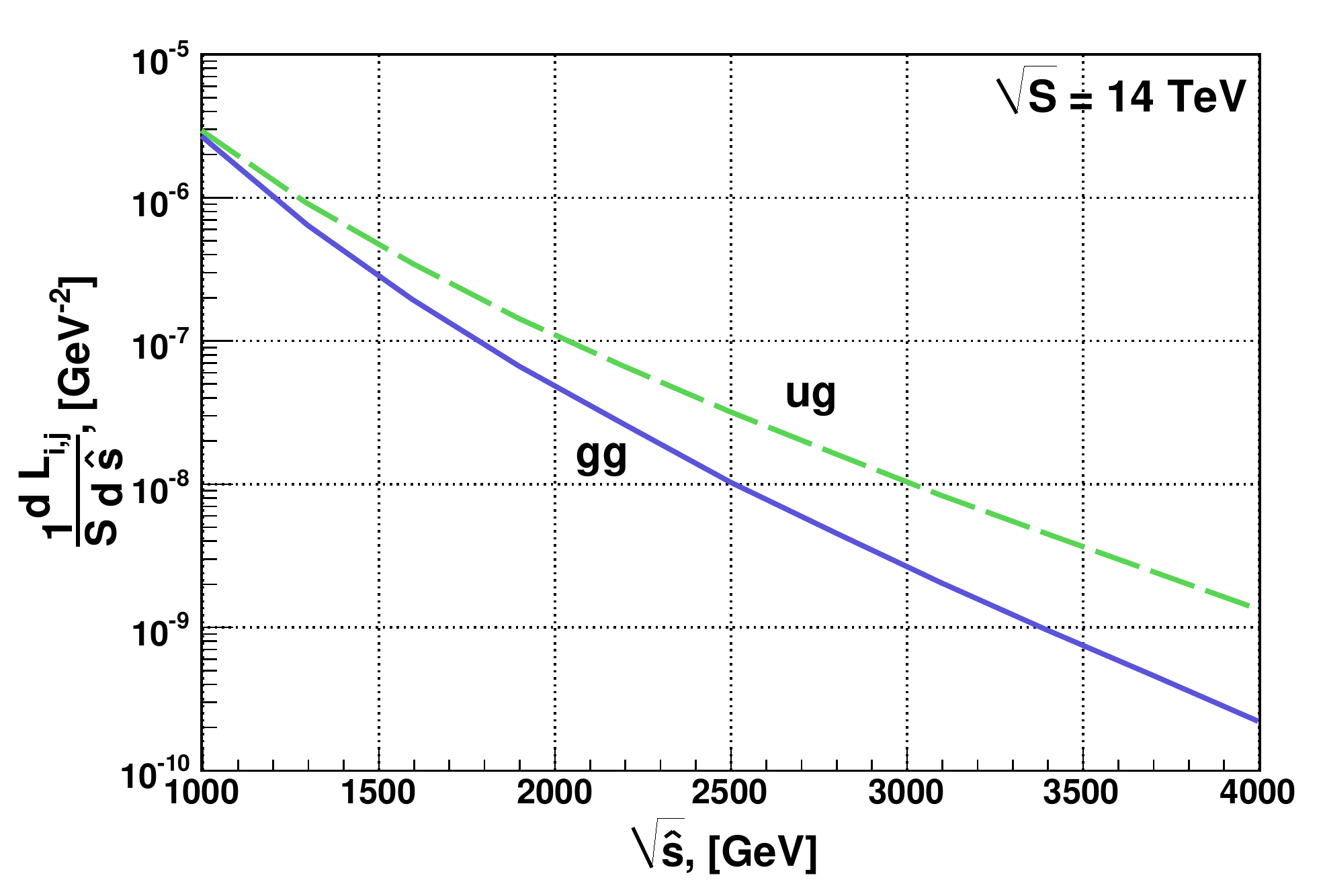}}} \\
	\end{tabular}
	\caption{On the left, the partonic cross-section for two typical contributions to single ($ug\rightarrow \T \bar t$; $\lambda=3$) and pair ($gg\rightarrow \T \bar{T}_{5/3}$) production.
		To find the total cross-sections, these have to be convoluted with the corresponding partonic luminosities which are shown on the right for $ug$ and $gg$ as function of $\sqrt{\widehat s}$.}
	\label{figPartonic}
\end{figure}

The main reason for the single production enhancement (or which is the same, for the pair production suppression) is explained by figure~\ref{figPartonic}, where the single and pair production 
partonic cross--sections are shown as a function of the partonic center--of--mass energy $\sqrt{\widehat{s}}=x_1x_2\sqrt{S}$ (where $\sqrt{S}=14$~TeV and 
$x_{1,2}$ the parton momenta fractions) for, 
respectively, $qg$ and $gg$ initial states and $1$~TeV top partner mass. 
Even though the the latter is bigger by a factor $\approx10$ in the first $500$ GeV $\sqrt{\widehat{s}}$ slice, it starts at higher $\sqrt{\widehat{s}}$ ($\sqrt{\widehat{s}}>2M$) while the threshold is lower ($\sqrt{\widehat{s}}>M+m_t$) in the single production case. The partonic cross--sections will 
have to be convoluted with the corresponding differential partonic luminosities which are defined as
$$
\displaystyle \frac{d{\mathcal L}_{i,j}}{d\widehat{s}}\,=\,\frac1S\int_{\widehat{s}/S}^1\frac{dx}x\,F_i(x)\,F_j(\widehat{s}/(Sx))\,,
$$
and shown in figure \ref{figPartonic}, computed using the MSTW PDF grids \cite{mstw} with $Q=2$ TeV. It is immediate to see that, since the differential luminosities decrease exponentially, the integrated one in the $[M+m_t,2M]$ range is much 
larger than the one from $2M$ and $\sqrt{S}$. The pair production total cross--section, which only receives contributions from the second  $\sqrt{\widehat{s}}$ 
interval ($\sqrt{\widehat{s}}\in [2M,\sqrt{S}]$), is suppressed w.r.t. single by a large factor. For the $1$~TeV case, the suppression factor is approximately given by 
${\mathcal L}_{i,j}(2M)/{\mathcal L}_{i,j}(M+m_t)\sim 1/20$, which is enough to compensate for the different partonic cross--sections of the two processes. 
Also, the suppression factor decreases for higher masses and this explains why the single production process is comparatively more important 
for higher masses, as figure~\ref{figXsec} shows.

Notice that, as discussed in \cite{Willenbrock:1986cr},  the single production process
can be considered as a $W$--gluon fusion because the intermediate $W$ tends (in order 
to maximize its propagator) to have 
low virtuality, of order $-m_w^2$, while still carrying enough energy to produce
the heavy partner. This makes the final state partonic line from which the $W$ is emitted to
be very forward, leading to a forward jet similar to the ones of $W$--boson
fusion Higgs production \cite{Cahn:1983ip}. For $p_T\lesssim m_w$, and an
energy around the TeV, the forward jet will have rapidity $\eta\gtrsim3$; the
presence of this forward jets constitutes an important feature of our signal
which we will discuss in more detail in the following.

\subsection{Signal and Background}

The one of same--sign dileptons is a very clean channel in which the single and
pair production of the top partner can be observed. The decay chains leading
to positively charged leptons for both the $B$ and $T_{5/3}$ production are
shown in fig.~\ref{figProdDiagram}, the case of negative charge final leptons
can be easily worked out. Notice that while the pair production process
contributes the same to both charges, single production leads to a charge
asymmetry. The hard initiating parton, which has to provide most of the energy
for  producing the partner, is indeed preferentially a valence quark and
specifically an up in the $(l^+,l^+)$ case and a down for $(l^-,l^-)$. We
therefore have a factor of roughly two between the single production
contribution to the positive and negative charge signal. This leads to a
sizable charge asymmetry that could be exploited to measure the single
production cross--section and eventually the top partner coupling, as we will
discuss.

Not to enter in subtleties of $\tau$ reconstruction, we only consider electrons
and muons and also, in order to neglect leptons from heavy flavors and
jet/lepton misidentification, we require leptons to be separated from hadronic
activity with a separation cut $\Delta R(LJ)>0.4$. Also, our leptons should be
hard enough ($p_T>10$~GeV) and of course inside the detector ($\eta<2.5$). Since
there are also neutrinos, our signal is in conclusion $pp\to l^{\pm} l^{\pm} +
\ET + n$ jets with hard and separated leptons. Notice that the separation cut,
in the case of very high mass, starts becoming costly for $\T$ since the leptons come
from boosted tops and tend to be close to the $b$--jet. 
This does not represent a problem for $\MT \lesssim 1.5$ TeV, but above it is a serious limitation.
But in this case, independently of this
effect, discovery in the dilepton channel will become hard due to the decrease of the cross--section worsened 
by the small leptonic $W$ branching ratio. In order to go further
hadronic $W$ final states should be included and advanced techniques of boosted
top reconstruction (see for example \cite{Agashe:2006hk-bulk}) will be needed.

The background, because of our isolation and hardness cuts,  comes from the
production of SM heavy particles subsequently decaying to leptons. Processes in
which exactly two same--sign leptons are produced are $W^\pm W^\pm$, $t\bar t
W^\pm$, $W^\pm W^\pm W^\mp$ and $t\bar t W^+ W^-$ where the latter two are
competitive with the former if the SM Higgs is heavy enough to give a resonant
contribution to the $W^+ W^-$ pair production. To maximize this possibility we
have assumed this to be the case and chosen the Higgs mass to be $m_H=180$ GeV.
Notice that we label the background processes only in terms of their heavy SM
particles content but we will take into account additional hadronic
activity they might also include. Strictly speaking, the suffix ``$+ n$ jets''
should be attached to all our background process names. There are also
processes in which three leptons are produced but one is lost, either outside
the detector ($\eta>2.5$) or below threshold, which we set at $5$ GeV. The only
relevant such a process is $W^\pm Z$ which, despite of the geometrical suppression factor
for loosing a lepton, is competitive with $W^\pm W^\pm$ since the latter final
state starts being produced, necessarily in association with at least two
partons, to a higher $\alpha_S$ order. The $W^\pm W^\pm Z$ process is on the
contrary subdominant compared with $W^\pm W^\pm W^\mp$ and can be neglected.

All the backgrounds listed up to now will be referred to as \textit{true
background}, while a second very important source of background, originating
because of charge misidentification, will be denoted as \textit{sign
background}. The latter are $t\bar t$, $Z^*/\gamma^*$ and $W^+W^-$ whose
cross--sections are clearly much larger than those of the true backgrounds we
have listed. Their real impact on our analysis depends on the charge
misidentification probability of the detector which is hard to estimate because
it strongly depends on the $\PT$ and rapidity of the leptons, as well as on the
detector and the lepton species (muons are typically better identified that
electrons, for instance). On top of this, it has been only poorly covered in
the literature the case of lepton momenta $\PT\sim 100-500$ GeV, which is the
region we are interested in. We will therefore use a flat charge
misidentification probability of $1\%$ believing this to be a conservative
assumption, rates closer to $10^{-3}$ (or even less for muons), being expected
\cite{tdr}.

\begin{table}[!t] 
\renewcommand{\arraystretch}{1.2}
\centering 
\begin{tabular}{|rl|r|r|} \hline &&
\multicolumn{1}{c|}{$\sigma(l^+l^+)$, [fb]}			& \multicolumn{1}{c|}{$\sigma(l^-l^-)$, [fb]}		\\ \hline 
\multicolumn{2}{|c|}{$\T$ + $\B$, $M = 0.5$ TeV}		&  84.6			&  45.2					\\ 
\multicolumn{2}{|c|}{$\T$ + $\B$, $M = 1.0$ TeV}		&  5.00			&  2.49					\\ 
\multicolumn{2}{|c|}{$\T$ + $\B$, $M = 1.5$ TeV}		&  0.596		&  0.272				\\ 
\multicolumn{2}{|c|}{$\T$ + $\B$, $M = 2.0$ TeV}		&  0.116		&  0.041				\\ \hline 
\multicolumn{2}{|c|}{$\T\T$ + $\B\B$, $M = 0.5$ TeV}		&  67.0			&  67.0					\\ 
\multicolumn{2}{|c|}{$\T\T$ + $\B\B$, $M = 1.0$ TeV}		&  1.47			&  1.47	 				\\
\multicolumn{2}{|c|}{$\T\T$ + $\B\B$, $M = 1.5$ TeV}		&  0.076		&  0.076 				\\ 
\multicolumn{2}{|c|}{$\T\T$ + $\B\B$, $M = 2.0$ TeV}		&  0.0053		&  0.0053				\\ \hline 
$t\bar t$		&+ 0, 1, 2 j				&  56.7			&  56.7					\\ 
$Z^*/\gamma*$		&+ 0, 1, 2, 3 j				&  168.0		&  168.0				\\ 
$W+W-$			&+ 0, 1, 2 j				&  5.83			&  5.83					\\ \hline 
$t\bar tW^\pm$		&+ 0, 1, 2 j				&  2.25			&  1.52					\\ 
$W^\pm Z$		&+ 0, 1, 2 j				&  7.66			&  4.44					\\ 
$W^\pm W^\pm$		&+ 2, 3 j				&  2.87			&  1.60					\\ 
$W^\pm W^\pm W^\mp$ 	&+ 0, 1, 2 j				&  1.97			&  1.28					\\ 
$t\bar tW^\pm W^\mp$	&+ 0, 1 j					&  0.595		&  0.595				\\ \hline 
\end{tabular} 
\caption{Cross-sections for the various
processes after a minimal set of cuts, $\PT(L)>10$ GeV, \mbox{$M(LL)>120$ GeV}; a
charge misidentification probability of $10^{-2}$ is taken into account. The
single production cross--section is for $\lambda_{T,B}=3$, by varying $\lambda$
it scales as $\lambda^2$.} 
\label{tabXsecMinCut} 
\end{table}

\subsubsection*{Simulation}

We generated signal and background events using \MGME  \cite{Alwall:2007st},
with higher order emissions of extra parton taken into account. \footnote{We
are indebted with R.~Contino for providing us with the \textsc{MadGraph} model
for the top partners.} Showering was performed with \Pythia 
\cite{Sjostrand:2006za}, where hadronization, which should not play any role in
our analysis, was turned off. For each process, we generated matrix element
events with different final state partons content and, after showering,
different matrix elements were combined by the MLM matching prescription with
$\kt$ jet algorithm \cite{Catani:2001cc, Alwall:2007fs, Mangano:2006rw}. Finally, jets were
reconstructed using the Paige's cone jet algorithm implemented in \getjet \cite{Soper:1986nv}, with parameters $\Delta R =
0.7$, $E_\textrm{min} = 20$ GeV. The resulting cross--sections are shown in
table \ref{tabXsecMinCut} where for each process the different matrix elements
combined by MLM matching are also reported. The
cross--sections of table~\ref{tabXsecMinCut} are for hard and separated leptons
inside the detector, as previously specified, but a cut $M(LL)>120$ GeV was
also implemented. The latter is needed to forbid real $Z$ production to
contribute to the Drell--Yan $Z^*/\gamma^*$ background. Also, the suppression
factor of $10^{-2}$ for charge misidentification is already taken into account.

The detailed simulation described above, including MLM matching, was performed
in order to obtain as realistic as possible samples of events, on which the
more detailed analysis of the top partners, which we will discuss in sect.~4,
can be reliably performed. The discovery results presented in this section,
however, do not rely on this detailed simulation. For event selection we will
indeed not use observables, such as for instance the number of jets, which
require a detailed knowledge of the hadronic structure of the event. We have
checked explicitly that the results of the present section can be reproduced by
pure matrix element simulations with additional hard QCD contributions.
This makes our analysis more robust.

In addition to the lepton charge misidentification, we have included the detector
 effect of fake $\ET$ \cite{tdr}.
A cut on $\ET$ is indeed very useful to get rid
of the $Z^*/\gamma^*$ background but fake $\ET$ is crucial for a realistic
estimate of the efficiency of this cut. Both for ATLAS and CMS the error on the
each component of the missing $p_T$ can be estimated by a Gaussian distribution
with a width given by the activity inside the detector: 
\begin{equation} \sigma
= \kappa \sqrt{\sum_{\textrm{jet,lep.}} |\PT|}\,, 
\end{equation} 
where for atlas $\kappa=0.46$ , while $\kappa=0.97$ for CMS \cite{tdr}.
We use $\kappa=1.0$ and, at the stage of
the analysis of the simulated events, add a random $\delta p_T$ fluctuation to
the missing $p_T$.

\subsection{Event Selection}

\begin{figure}[!p]
	\begin{center}
	\begin{tabular}{cc}
		\scalebox{0.35}{\rotatebox{90}{\includegraphics{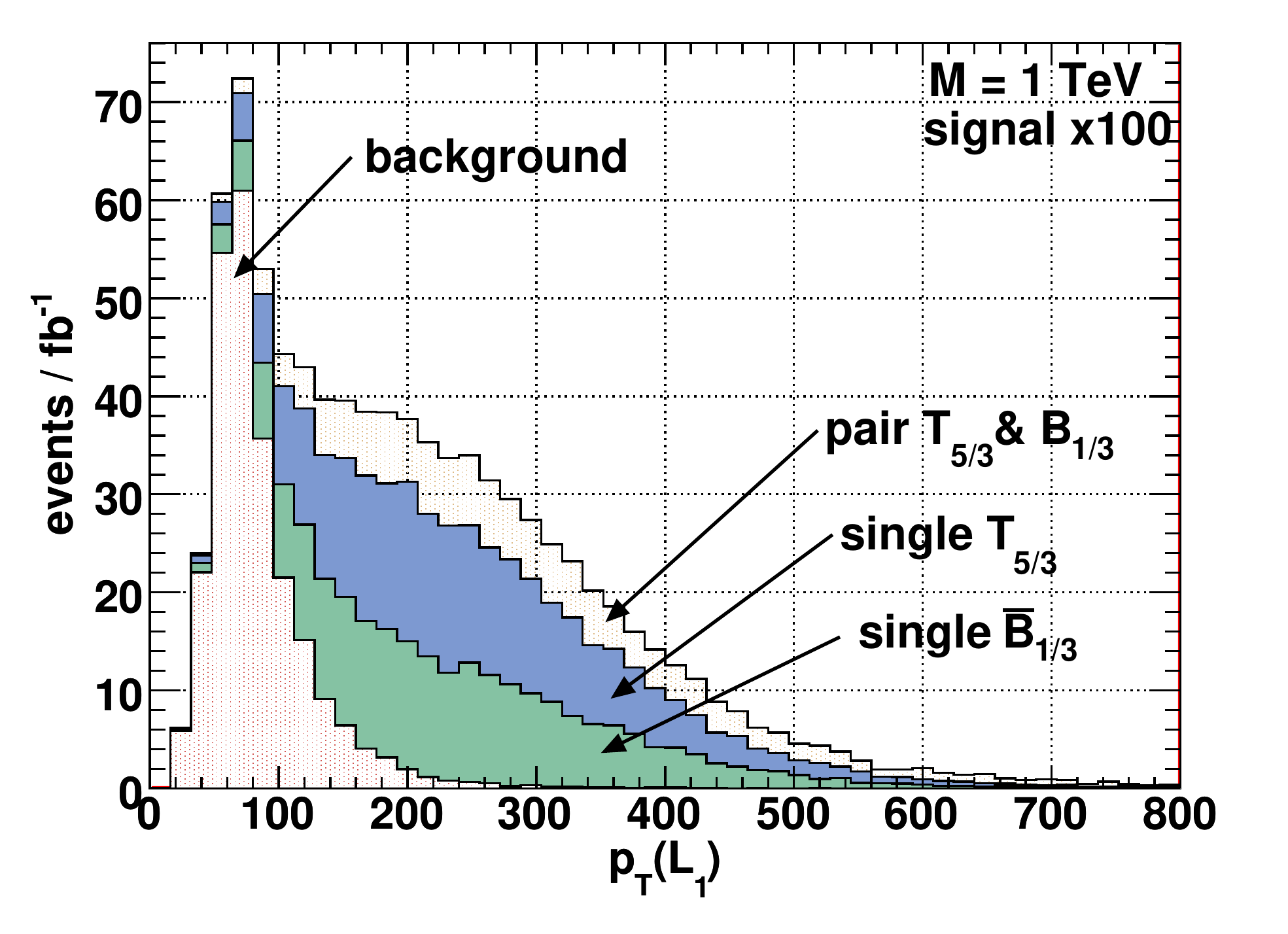}}} & 
		\scalebox{0.35}{\rotatebox{90}{\includegraphics{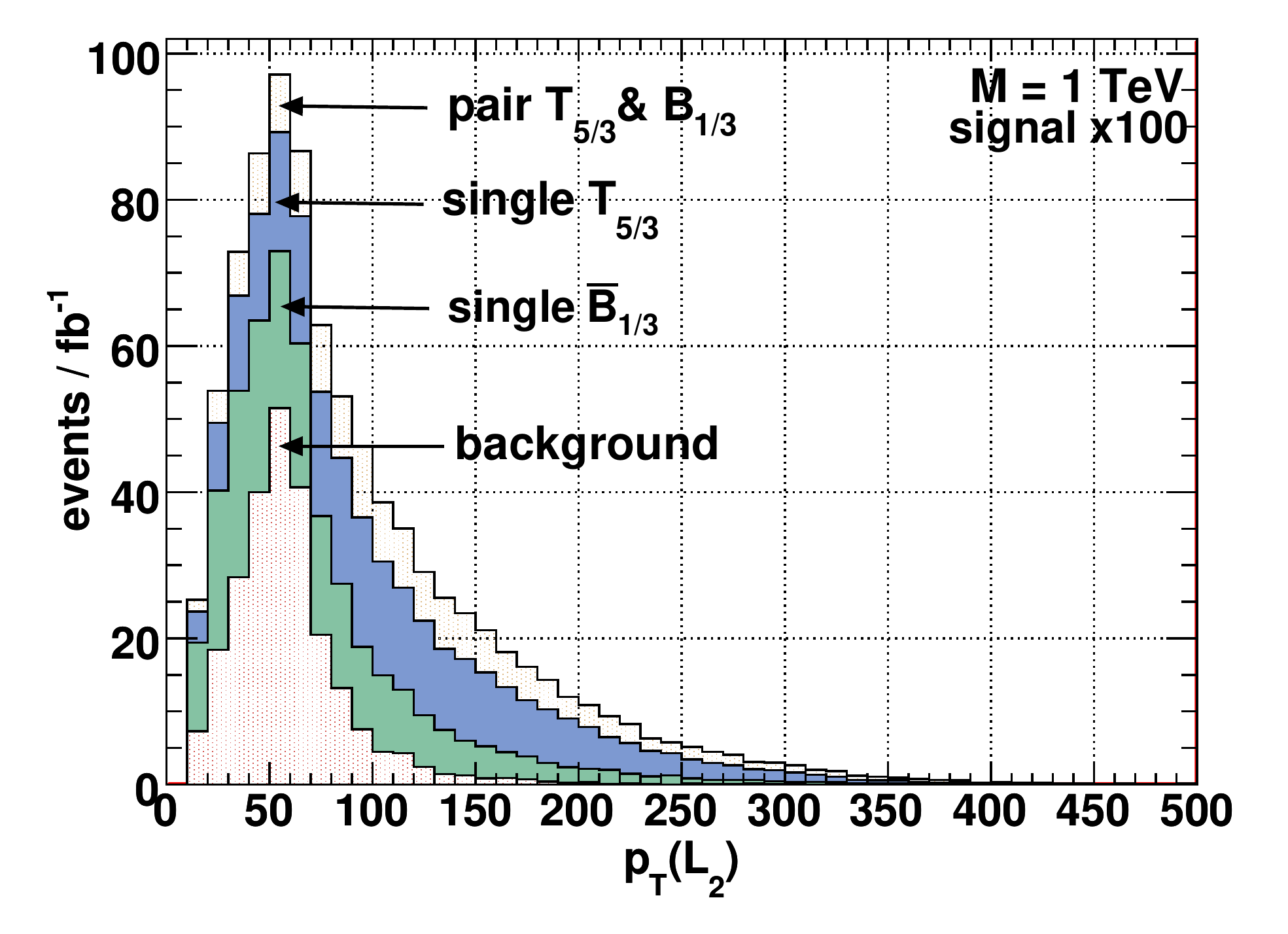}}} \\
		\scalebox{0.35}{\rotatebox{90}{\includegraphics{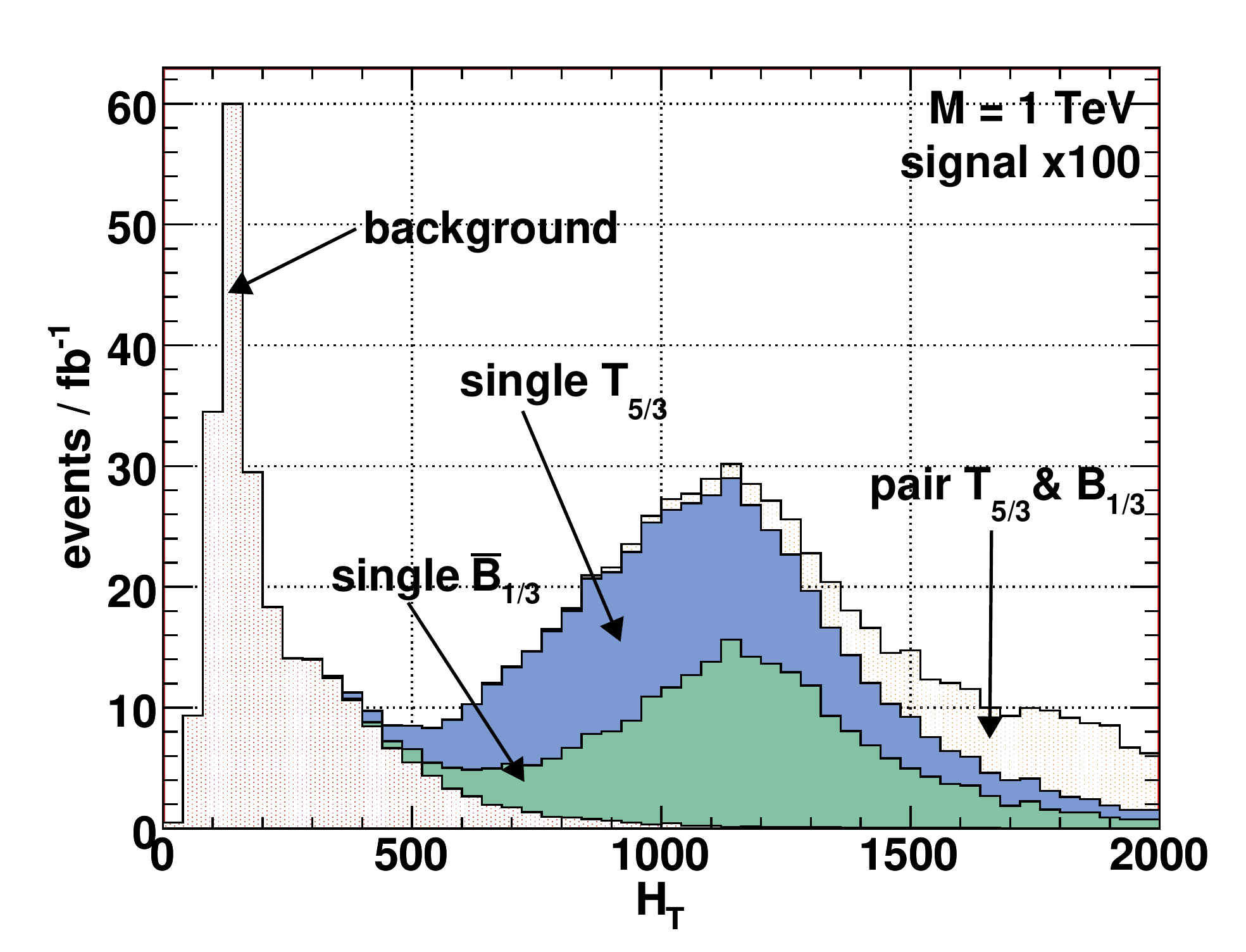}}} &
		\scalebox{0.35}{\rotatebox{90}{\includegraphics{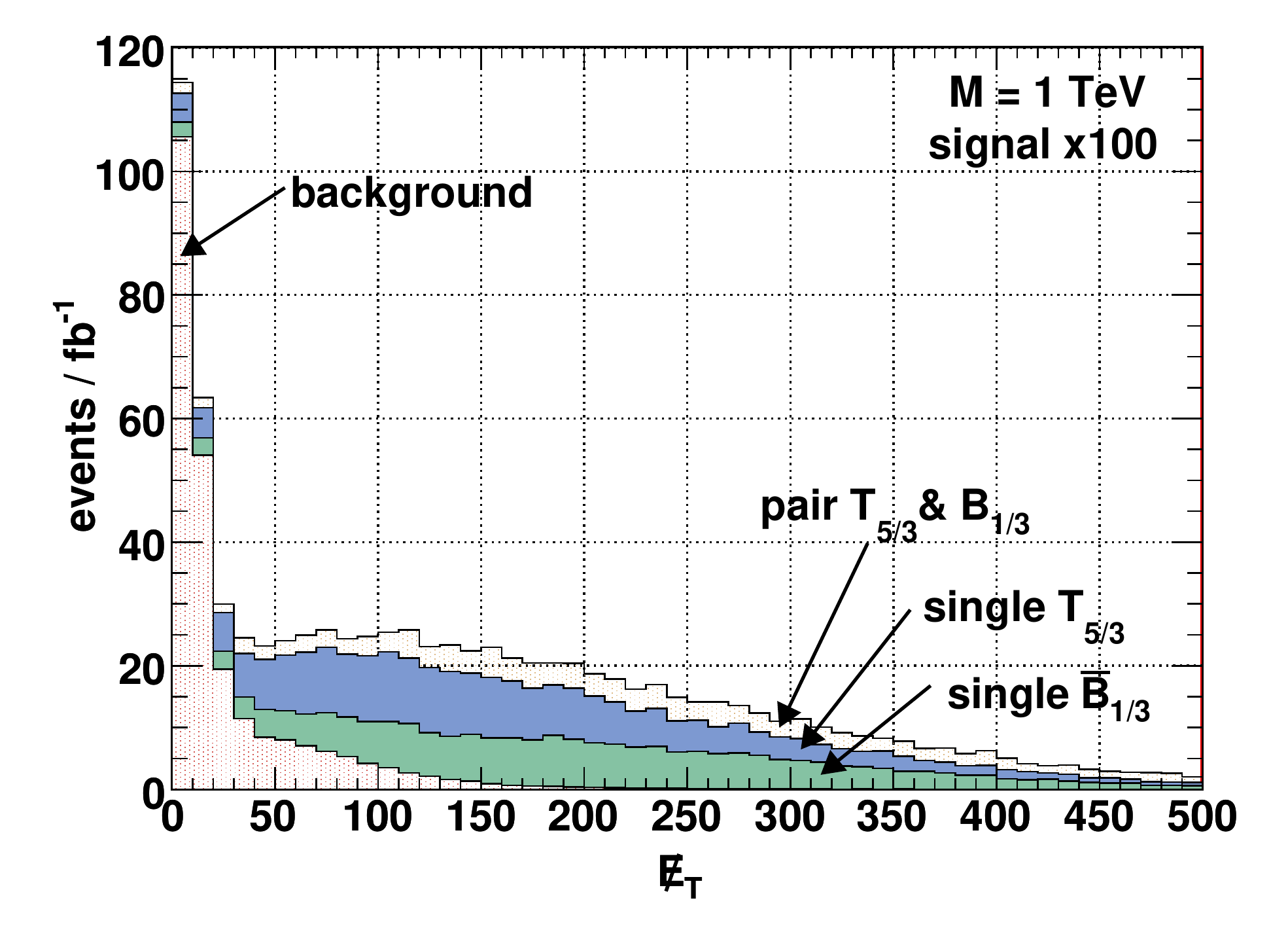}}}
	\end{tabular}
	\end{center}
	\caption{Distribution of $\PT(L_1)$, $\PT(L_2)$, $H_T$ and $\ET$ for signal ($\times 100$) and background after the minimal set of cuts, $\PT(L_{1,2}>10)$ GeV, $M(LL)>120$ GeV. The case $\lambda_{T,B}=3$ is considered.}
	\label{figDistribMinCut}
\end{figure}

The dominating background, as table~\ref{tabXsecMinCut} shows, is $Z^*/\gamma^*$, but this will become substantially irrelevant when a cut on $\ET$ will be applied. Second comes $t\bar t$ which will be more difficult to get rid of. The choice of the observables and their optimization focuses on the single production, which is the relevant production mechanism in the range of high masses where discovery becomes less easy. Our search strategy, however, turns out to be efficient for pair production as well. The first observables we focus on are the $\PT$ of the hardest and second hardest leptons, $L_1$ and $L_2$, whose distributions are shown figure \ref{figDistribMinCut}. Notice that the leptons are typically softer for $B$ than for $T_{5/3}$, but this had to be expected since in the second case both leptons come from the heavy particle decay while in the first only one comes from the $B$ and the other originates from the top quark. A cut on $p_T(L_1)$ will nevertheless be useful also for the $B$, but pushing it to too high values would induce a big unbalance between $T_{5/3}$ and $B$ lowering too much the cross--section for the latter. The $\PT$ of the second lepton, which would not in any case solve this problem, turns out to be an inefficient cut because a hard cut on $p_T(L_1)$ already implies, on the background, a certain hardness of $p_T(L_2)$. This is the case, for instance, for the dominant $t\bar t$ background, we therefore find it convenient to keep the cut on $p_T(L_2)$ to its minimal value of $10$ GeV.

\begin{table}[!p]
	\centering
	\begin{tabular}{|cc|r|r|r|}\hline
	Cut	& Mass, [TeV]	& $\PT(L1)$, [GeV]	& $\HT$, [GeV]		& $\ET$, [GeV]	\\ \hline
	soft	& 0.5		& 60			& 500  			& 50		\\ \hline
	medium	& 1.0		& 100			& 1000			& 50		\\ \hline
	hard	& 1.5		& 200			& 1200			& 100		\\ \hline
	max	& 2.0		& 250			& 1600			& 100		\\ \hline
	\end{tabular}
	\caption{Table of the cuts depending on the mass of the partners.
	To these, there is a cut on the invariant mass of both leptons, $M(LL)>120$ GeV, and the $\PT$ of the second lepton, $\PT(L2)>10$ GeV.}
	\label{tabCuts}
	\vspace{0.8cm}
	\begin{tabular}{|c|r|r|r|r|r|r|r|r|}\hline
			& \multicolumn{2}{c|}{soft}& \multicolumn{2}{c|}{medium}& \multicolumn{2}{c|}{hard}& \multicolumn{2}{c|}{max}		\\ \hline
	Process		& ++	& -{}-	& ++		& -{}-		& ++	& -{}-	& ++		& -{}-					\\ \hline
	Single $\T$	& 38.4	& 18.7	&   1.50	& 0.64		& 0.154	& 0.059	& 0.015		& 0.005					\\
	Pair $\T$	& 28.1	&   28.1&  0.662	&  0.662	&   0.03&   0.03& 0.0021	& 0.0022				\\ \hline
	Single $\B$	& 32.4	& 14.4	&   1.76	& 0.72		& 0.226	& 0.085	& 0.042		& 0.015					\\
	Pair $\B$	&   26.3&   26.3&  0.651	&  0.651	&   0.03&   0.03& 0.0022	& 0.0022				\\ \hline
	True background	& 3.5	& 2.0	&   0.68	& 0.34		& 0.174	& 0.077	& 0.058		& 0.020					\\ \hline
	Sign background	& 12.3	& 12.3 	&   0.72	& 0.72		& 0.100	& 0.100	& 0.009		& 0.009					\\ \hline
	\end{tabular}

	\caption{Cross-section in [fb] for the different processes.
	For the signal, the mass corresponding to the cut is used.}
	\label{tabXsecAfterCuts}
\end{table}

The essence of our signal is simply the decay of a heavy particle, whose energy needs to be distributed among its different products, leptons, jets ($J$) and $\ET$. It is therefore useful to define the total transverse energy $\HT$:
\begin{equation}
	\HT = \sum_{J,L,\ET} |\vec\PT|\,.
\end{equation}
The $\HT$ distribution is shown in figure~\ref{figDistribMinCut} and, as expected, peaks around the heavy particle mass. A cut on $\HT$ will therefore be extremely useful and will also solve the problem of the unbalance among $B$ and $T_{5/3}$ due to the $\PT(L_{1})$ cut since $\HT$ is typically bigger for $B$ than for $T_{5/3}$. The reason why $\HT$ is bigger for $B$ is that $\ET$, which directly contributes to $\ET$, is also bigger and this is due to the following. The partner has, when singly produced, low transverse boost so that the $T_{5/3}$ decay products, the $t$ and the $W$, are back to back in the transverse plane and boosted. This favors the two neutrinos from $t$ and $W$ to be back to back and in this configuration there is a cancellation in $\ET$. This cannot happen for the $B$ where there is only one hard neutrino. It will nevertheless be hard, for masses $>1.0$ TeV, to balance the $\B$ and $\T$ signals after cuts. The lepton-jet separation will indeed unavoidably disfavor the $\T$ selection since it forbids too boosted leptonic tops. 

The two main cuts we will use are, in conclusion, $p_T(L_1)$ and $\HT$, but we will also ask some $\ET$ and $M(LL)>120$ GeV as mentioned above to get rid of the $Z^*/\gamma*$ background. For different masses, we optimize our cuts for the case in which both the $T_{5/3}$ and $B$ are present, have degenerate masses and ``average'' coupling $\lambda_{T,B}=3$. The optimized values are shown in table \ref{tabCuts} and the corresponding cross--sections are reported in table~\ref{tabXsecAfterCuts}.  We define the discovery luminosity to be the one for which $S/\sqrt{B}=5$ or, if the background is negligible as in the $0.5$~TeV mass case, as the luminosity which is needed to observe $5$ signal events; the results are shown in table~\ref{tabLdisc}. We see that discovery will be possible, in the degenerate mass case, up to at least $1.5$~TeV top partner mass while it appears difficult for $2$~TeV even when the entire LHC program of $300$~fb$^{-1}$ total luminosity will be completed. For masses below around $1$~TeV, moreover, even $100$~fb$^{-1}$ of luminosity will allow to collect a significant number (greater than $800$) of signal events which will allow to study the top partners in some detail measuring their masses and couplings as we will discuss in the following section.

The situation is worst, clearly, when only the $B$ partner is present. The discovery luminosity for different masses and for $\lambda_{B}=3$ are shown in table~\ref{tabLdisc1} after applying the cuts of table~\ref{tabCuts}. Though the cuts are not optimized for this case, we see that the discovery is impossible for  $2$~TeV while the case of $1.5$~TeV is within the reach of the LHC. Lowering the coupling renders the discovery more difficult for high mass, where the pair production is small, given that the single production cross--section decreases as $\lambda^2$. For $\lambda_{T,B}=2$ and degenerate masses, for instance, $90$~fb$^{-1}$ are needed for $1.5$~TeV mass while the $2$~TeV case is by far beyond reach. When only the $B$ is present and $\lambda_{B}=2$, the $1.0$ case should be discovered with $11$ fb$^{-1}$, while the  $1.5$~TeV case becomes difficult as the discovery luminosity, though again estimated with the cuts of table~\ref{tabCuts} which are not optimized for this case, is of $287$~fb$^{-1}$.

\begin{table}[!t]
\centering
\begin{tabular}{|c|c|c|c|}\hline
	Mass, [TeV]	& $L_\textrm{discovery}$, [fb$^{-1}$]	& $\#$ signal & $\#$ background \\ \hline
	0.5	& 0.024		& 5	& 0	\\ \hline
	1.0	& 1.103		& 8	& 2	\\ \hline
	1.5	& 26.40		& 17	& 11	\\ \hline
	2.0	& 326.7		& 28	& 31	\\ \hline
\end{tabular}
\caption{Discovery luminosity, for different mass, in the case of degenerate top partners with $\lambda_{T,B}=3$.}
\label{tabLdisc}
\vspace{0.5cm}
\begin{tabular}{|c|c|c|c|}\hline
	Mass, [TeV]	& $L_\textrm{discovery}$, [fb$^{-1}$]	& $\#$ signal & $\#$ background \\ \hline
	0.5	& 0.076		& 8	& 2	\\ \hline
	1.0	& 4.3		         & 16	& 11	\\ \hline
	1.5	& 82   		& 30	& 37	\\ \hline
	2.0	& 637		& 39	& 61	\\ \hline
\end{tabular}
\caption{Discovery luminosity, for different mass, when only the $B$ partner is present and $\lambda_{B}=3$.}
\label{tabLdisc1}
\end{table}

The top partners, even if both exist, need not to be degenerate or to have equal couplings. Actually, an unavoidable source of splitting is that, as discussed in sect.~2, the $B$ mixes with the $b_L$ while the $T_{5/3}$ does not. In the specific model described in sect.~2 this results in $M_B= M_{Q}/\cos{\varphi_q}$ while $M_T=M_Q$ which implies:
\begin{equation}
\displaystyle{\frac{M_T}{M_B}\,=\,\frac{\lambda_B}{\lambda_T}\,=\,\cos{\varphi_q}}\,,
\label{mcoup}
\end{equation}
where $\varphi_q$ is the mixing angle of the $q_L$ doublet. 
Since the $q_L$ will not be very composite to satisfy experimental constraints, {\it{i.e.}}  $\sin{\varphi_q}$ is small, in our model the partners are likely to have similar masses, though the $B$ will always be heavier than the $T_{5/3}$ and also more weakly coupled since $\lambda_B<\lambda_T$. The $B$ cross--section will therefore quickly decrease, especially for high masses where single production is more relevant, by increasing $\sin{\varphi_q}$, and a valid (although pessimistic) approximation of this case is to neglect the entire $B$ partner contribution. This leads us to consider the case in which only the $T_{5/3}$ is present and $\lambda_T=2$, which correspond to maximal $\sin{\varphi_q}$. We obtain that the discovery is surely possible for $1$~TeV mass, with $10$~fb$^{-1}$, but it is very difficult for $1.5$~TeV where $470$~fb$^{-1}$ would be needed. This case is worst than the one of the $B$ since our cuts are, as we have explained, more efficient for the $B$ than for the $T_{5/3}$ if $\MT > 1.0$ TeV. Even though discovery is difficult in this unfavorable situation for exactly $1.5$~TeV mass, the cross--section decreases so fast 
with the mass that already for $1.3$~TeV discovery is possible, $90$~fb$^{-1}$ being required.
Intermediate cases obeying eq.~\ref{mcoup} could also be considered, but we prefer not to restrict to the specific model of sect.~2 and instead consider more general situations. Similar but not equal top partner masses and couplings will be considered in the following section where we will discuss the LHC phenomenology of the top partners in more detail.

Our event selection strategy is also efficient for the pair production, as we have mentioned. If only pair production is considered, indeed, we find (when both partners are present) a discovery luminosity of $64$~pb$^{-1}$ and of $8.9$~fb$^{-1}$ for, respectively, $0.5$ and $1.0$~TeV mass, and this has to be compared with the values of $56$~pb$^{-1}$ and of $15$~fb$^{-1}$ which have been found in \cite{Contino:2008hi}. Since our background is larger, especially for low mass, due to the charge misidentification effect which was ignored in \cite{Contino:2008hi}, our strategy seems more efficient.

\section{Phenomenology of the top partners}

The discovery of an excess in the  same sign dilepton channel would not be a proof of the existence of the top partners since other new physics scenarios, or an erroneous estimate of the background cross--section, could lead to the same effect. The aim of this section is to underline the main features of the top partner signal, which will be crucial for recognizing it, and also to propose some strategies for measuring the partner's masses and couplings. Since a large number of events is required to perform reliable measures, the results presented in this section are expected to be useful for masses below $1.2$ or $1.3$~TeV, the cross--section being likely too be too small in the $1.5$~TeV case.

\subsection{SM particles}

Let us first of all try to identify the heavy SM particles, the $W$'s and the hadronic top, which are present in our signal. This will already give us some confidence in the hypothesis that the dilepton excess is due to a modified charged current in the top sector, even though this feature is shared by the $ttW$ and $ttWW$ backgrounds. Identifying the $t$ will also help us later to reconstruct the new particles which would constitute the most striking evidence for their existence.

\begin{figure}[!tb]
	\begin{center}
		\scalebox{0.6}{\includegraphics{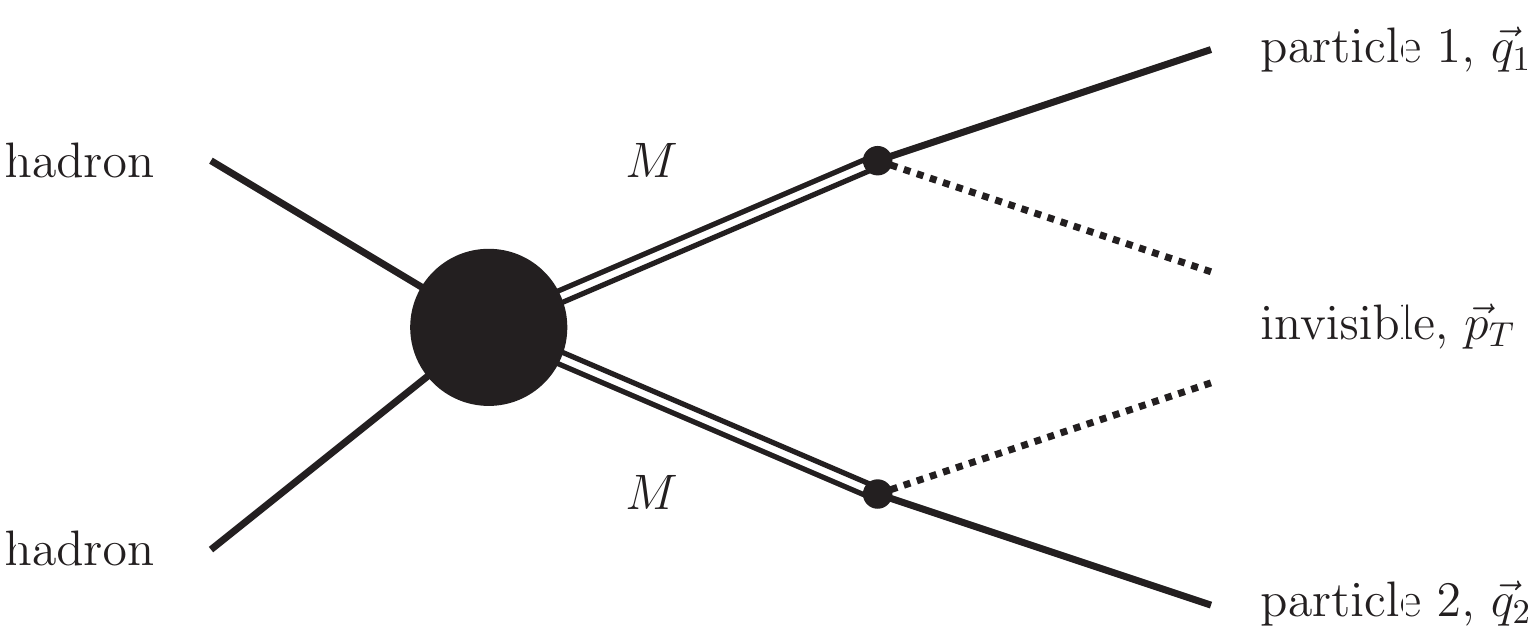}}
	\end{center}
	\caption{Topology for which $\mtt$ is defined: pair production of particle with semi-invisible decay}
	\label{figMt2Topology}
\end{figure}

\begin{figure}[!t]
	\centering
	\begin{tabular}{cc}
		\parbox[c]{8cm}{\rotatebox{90}{\scalebox{0.4}{\includegraphics{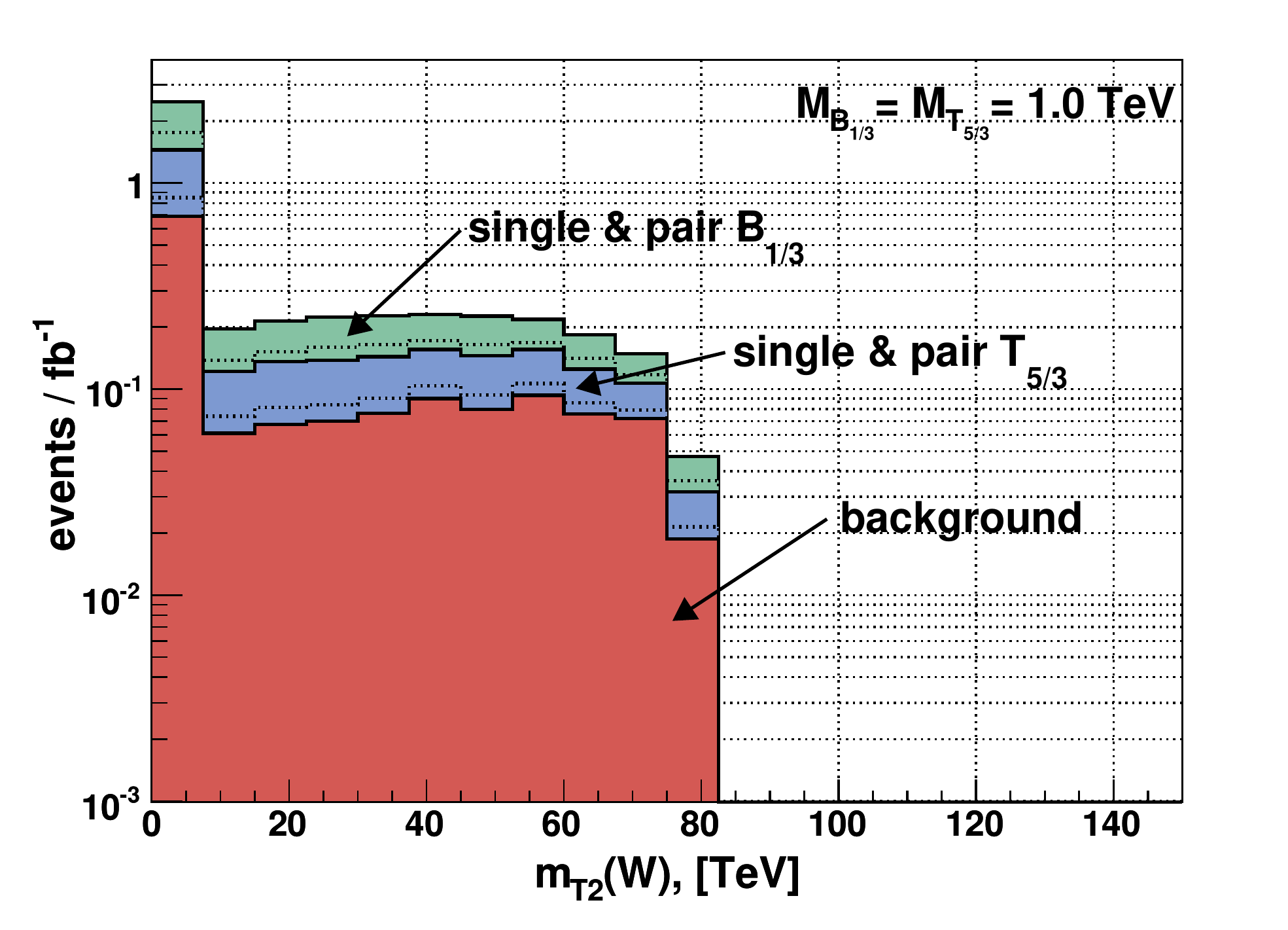}}}} &
		\parbox[c]{3cm}{\scalebox{0.6}{\includegraphics{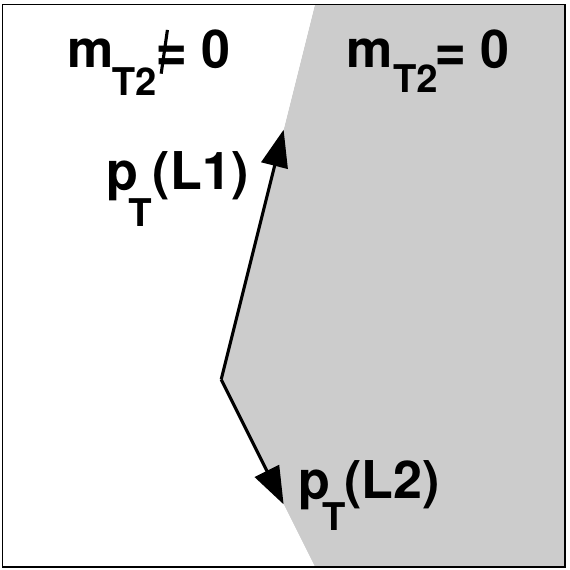}}}
		\end{tabular}
	\caption{On the left, $\mtt$ distribution for the signal with $\MT=\MB=1.0$ TeV, $\lambda_{T,B}=3$ selecting only the $l^+l^+$ final state.
		The large peak at $\mtt=0$ is due to configurations in which the lepton and neutrino transverse momenta could be parallel, which is the case when the $\ET$ vector falls in the red region shown on the right.}
	\label{figMt2W}
\vspace{0.8cm}
	\begin{tabular}{cc}
		\rotatebox{90}{\scalebox{0.25}{\includegraphics{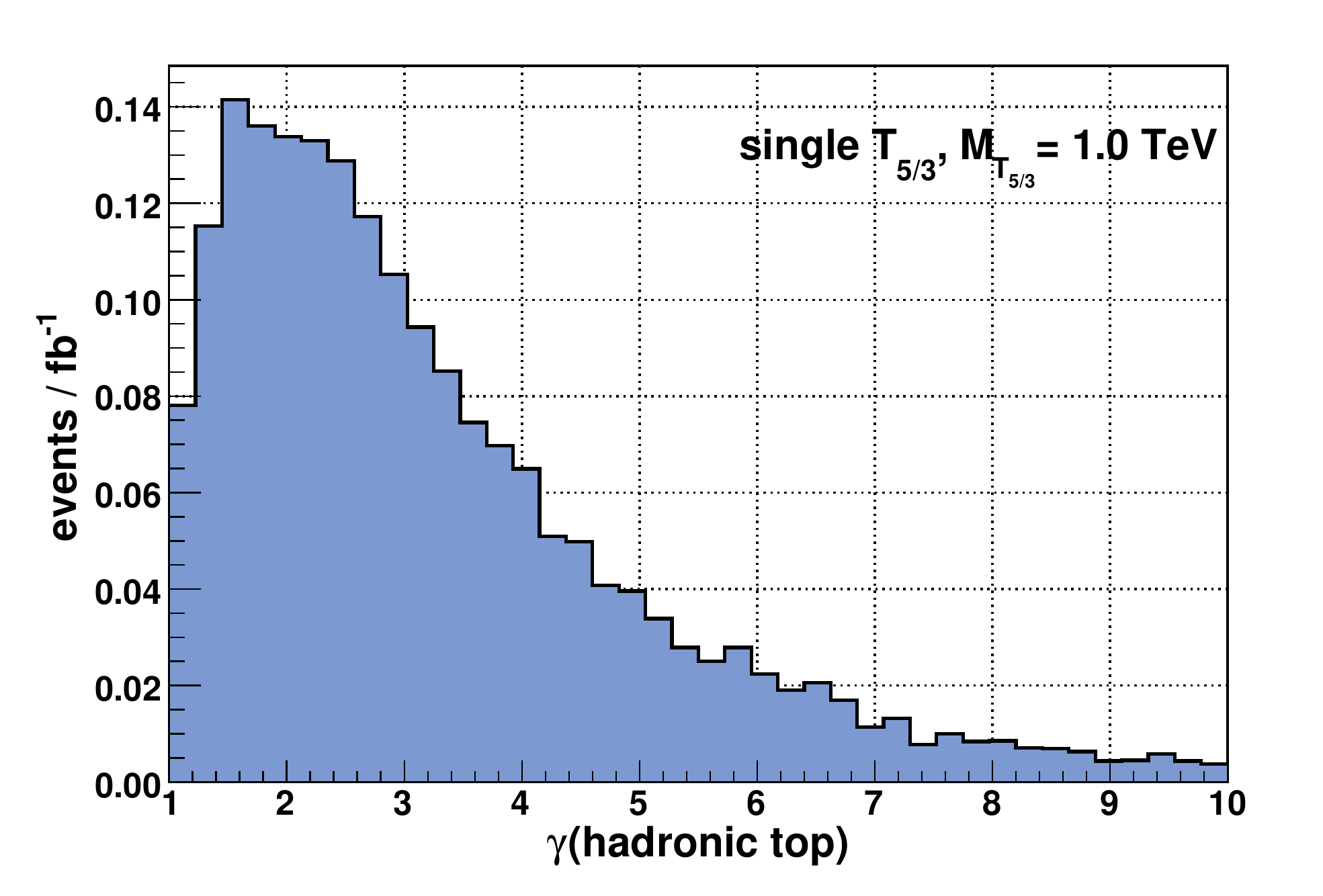}}} &
		\rotatebox{90}{\scalebox{0.25}{\includegraphics{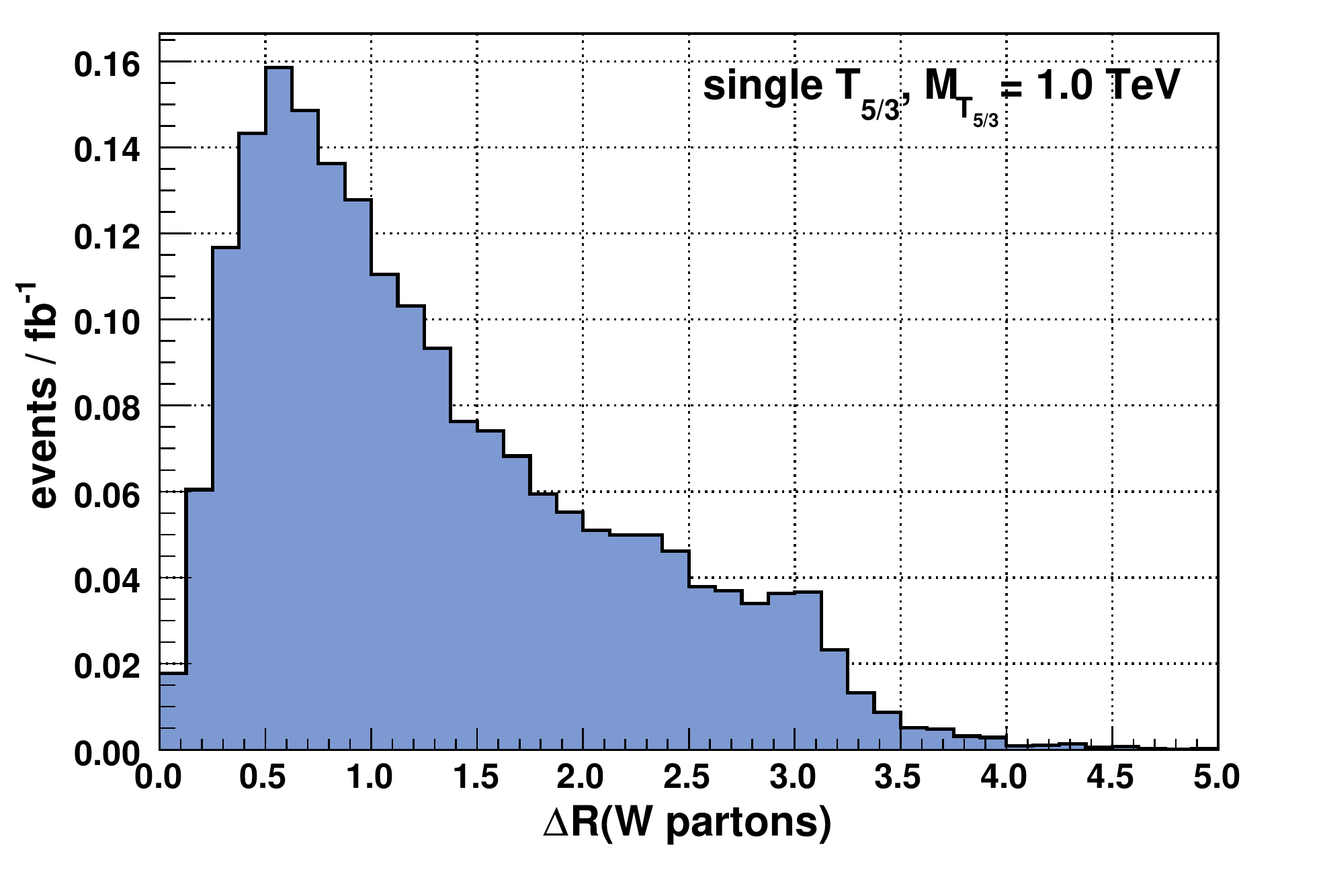}}} \\
		\rotatebox{90}{\scalebox{0.25}{\includegraphics{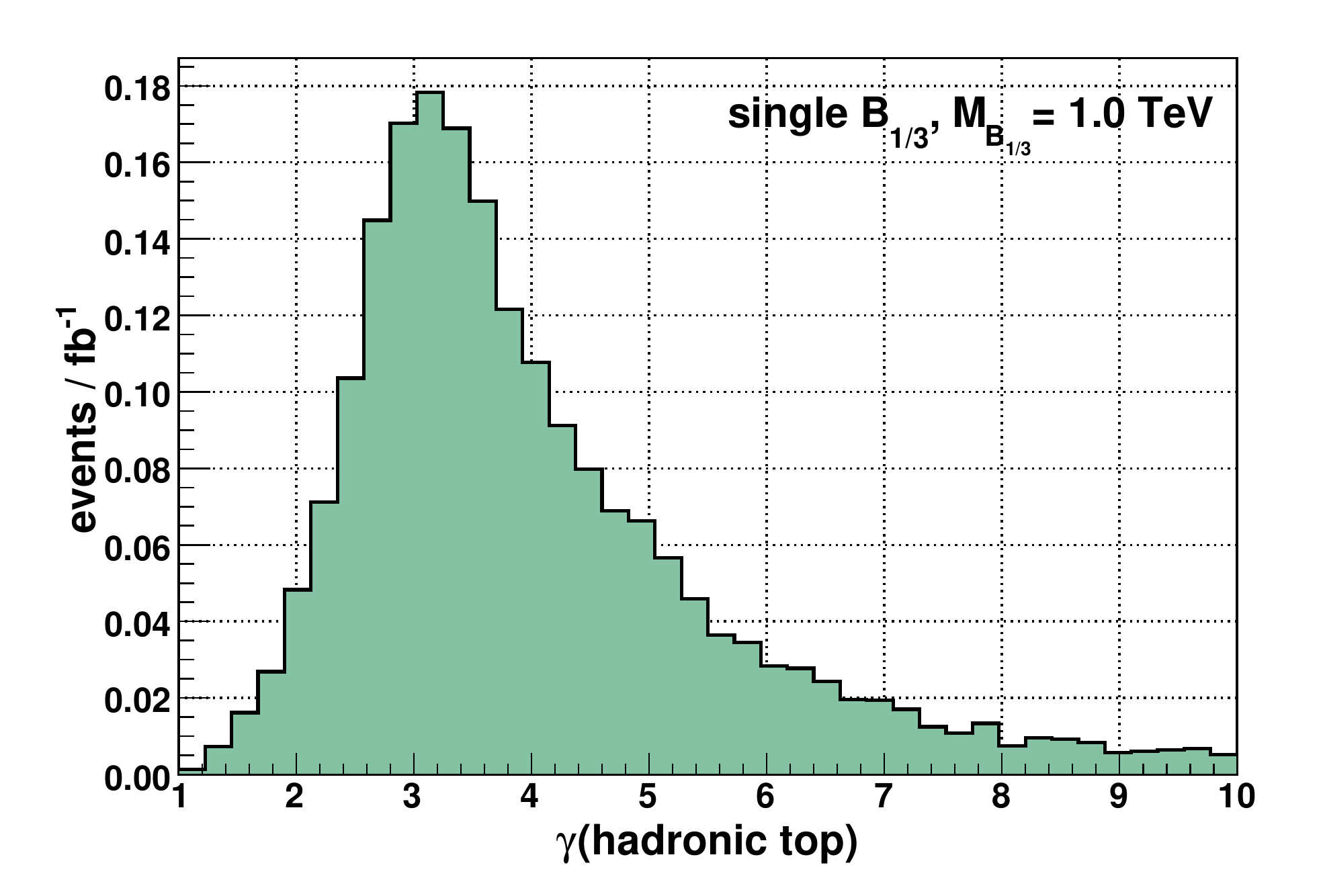}}} &
		\rotatebox{90}{\scalebox{0.25}{\includegraphics{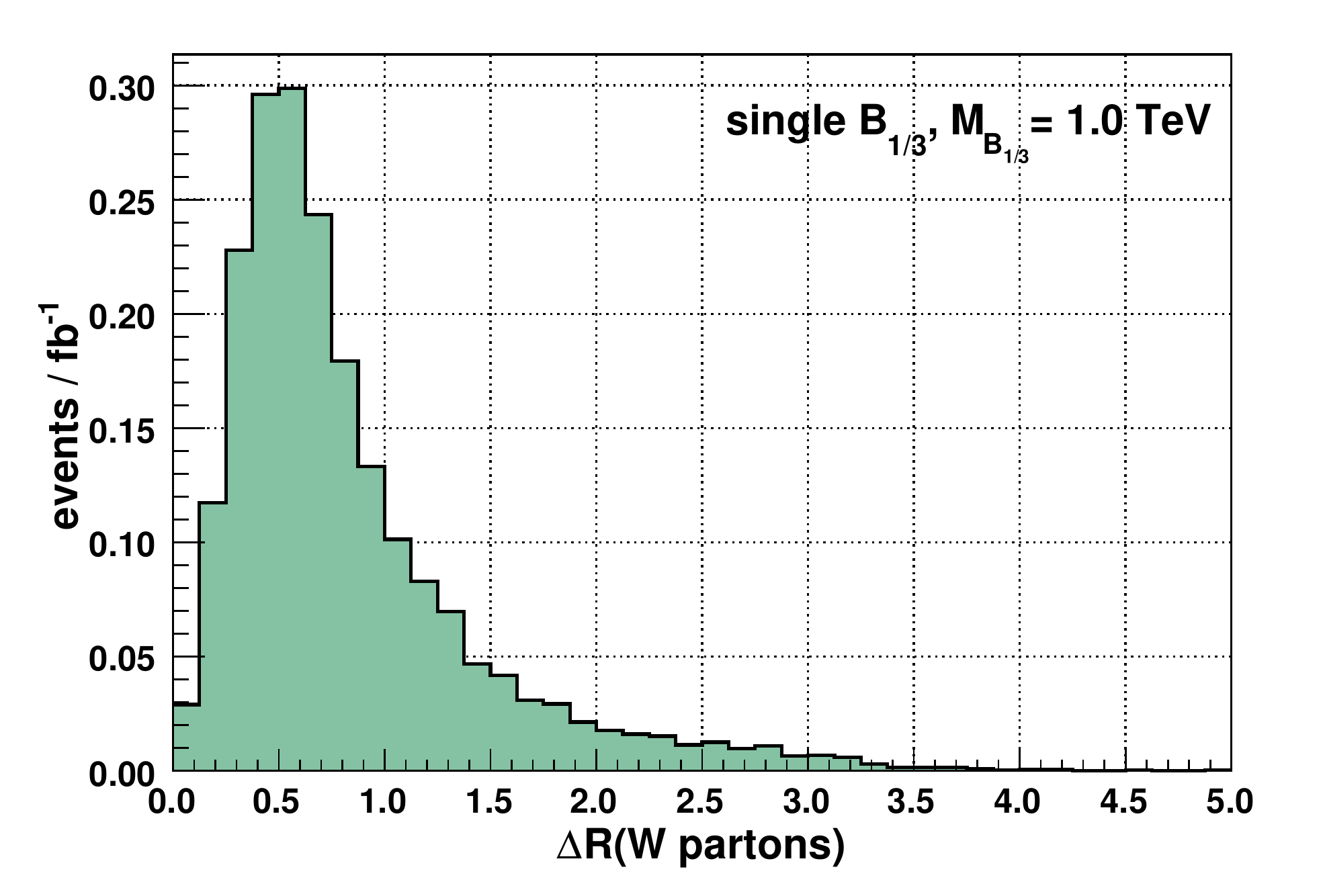}}} \\
	\end{tabular}
	\caption{The left plots show the Lorentz factor $\gamma$ of the hadronic top for $\T$ and $B$, while the right plots shows the $\Delta R$ separation 	of the partons origination from the $W$ decay. Signal events are used with $M_{T,B}=1.0$~TeV, $\lambda_{T,B}=3$ and ``medium'' cuts are applied.
		}
	\label{figHadTop}
\end{figure}

In our signal the leptons come from the decay of $W$ bosons and even if this property is shared by most of the backgrounds, and is quite common for leptons in a hadronic environment, it is good to establish this fact in order to exclude, for instance, an exotic particle coupled to the charged lepton current such as a heavy $W'$. The presence of the $W$'s in our signal, where the only sources of $\ET$ are the two neutrinos, is easily observable by the end--point of the  $\mtt$ (also called stransverse mass) \cite{Lester:1999tx-bulk} distribution. This variable is designed to extract the mass $M$ of a pair produced particle with semi--invisible decay as shown figure \ref{figMt2Topology}. It is defined as
\begin{equation}
\label{eqMt2}
	\mtt = 	\min_{ \vec{\slashed{p}}_T^1 + \vec{\slashed{p}}_T^2 = \vec{\slashed{p}}_T }\left\{ 
			\max_{i=1,2}\left\{ 
				\mt( \vec{\slashed{p}}\mbox{}_T^{~i}, \vec{q}\mbox{}_T^{~i})
			\right\}
		\right\}
		\leq M\,,
\end{equation}
where $\vec{\slashed{p}}_T$ is the $\ET$ vector and $\vec{q}\mbox{}_T^{~i}$ are the transverse components of the visible part of the decay. In our case, $\vec{q}\mbox{}_T^{~i}$ are the leptons transverse momenta and the $\mtt$ distribution ends at $m_W$ as shown in figure~\ref{figMt2W}. It should be noted that when, as in our case, the transverse mass $\mt$ in eq.~(\ref{eqMt2}) is computed with massless particles (the lepton and the neutrino), $\mtt$ will vanish if the neutrino momenta ${\slashed{\vec{p}}}_T^{1,2}$ can be chosen to be parallel and aligned with the leptons. This is always possible when the $\ET$ vector lays inside the angle between the lepton transverse momenta and, since our leptons are usually back to back, this will roughly happen one half of the times. The $\mtt$ distribution therefore shows a peak at zero and, since the events in the peak are useless for determining the threshold, this effects reduces the efficiency of the procedure by a factor one half.

The signal also contains a hadronic top which we can reconstruct in two steps. First we reconstruct the hadronic $W$ and afterwards we associate to it the corresponding jet from the top decay. The latter will be a $b$ jet, but $b$--tagging will not be needed. Except for single production of $\T$ , the top will be slightly boosted since it comes from the decay of a heavy particle, as shown in figure~\ref{figHadTop}. The $W$ and the $b$ will still be separated enough to be resolved while the two $W$ jets have a certain probability (we use a cone jet with $\Delta R = 0.7$ and $\ET_{min}=20$ GeV) to merge into a single jet, as figure~\ref{figHadTop} shows. We therefore proceed as follows. First we look for a single jet with $|m(J)-m_W| \leq 20$ GeV and if none is found, we do the same with all the jet pairs $|m(J_1+J_2)-m_W| \leq 20$ GeV. To each $W$ reconstructed in this way we try to associate a jet such that $|m(W+J)-m_t|\leq 30$ GeV and when this is achieved the top is considered to be reconstructed. For the signal, we estimate the efficiency of this procedure to be above $60\%$, even though a detailed detector simulation would be needed for a more detailed identification and hence a realistic estimate.

\subsection{Evidences of single production}

If the top partners coupling to top is large, as theoretically expected, a sizable fraction (or even the entire sample, for high mass) of the dilepton events originates from single production and establishing this fact will permit to distinguish the top partner from a generic pair--produced new heavy colored fermion. Studying single production will also allow to measure the top partners coupling.

\begin{figure}[!tb]
	\centering
	\rotatebox{90}{\scalebox{0.35}{\includegraphics{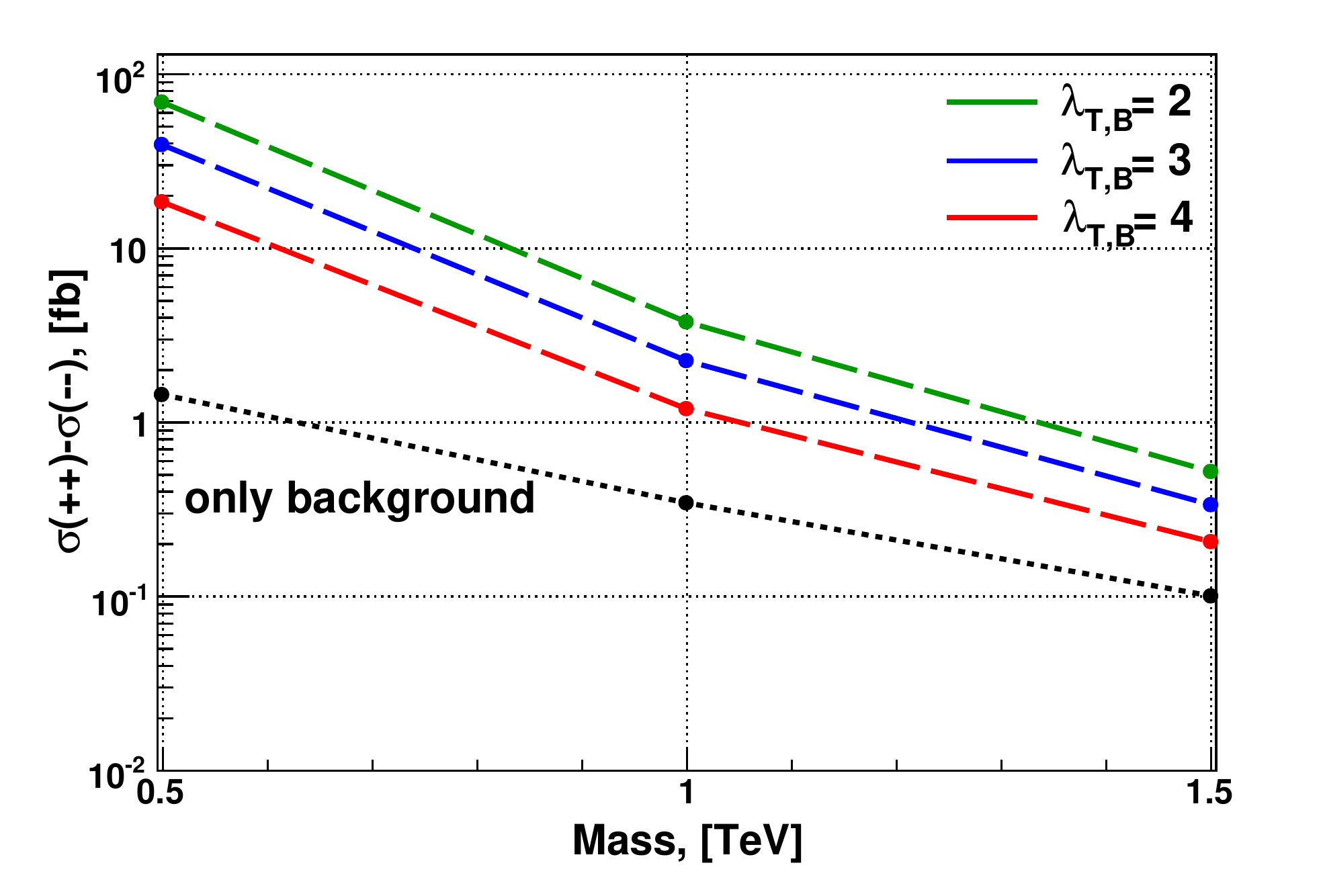}}}
	\caption{The cross--section difference $\sigma(++)-\sigma(--)$, at different masses, of signal + background for different values of $\lambda_{T,B}=2,3,4$.
		For each mass, the corresponding set of cuts is used.}
	\label{figChargeAsymmetry}
\end{figure}

The first evidence of single production is the presence of a charge asymmetry which is due, as discussed in section~3, to the fact that the single production originates from a $W$ radiated by an initial quark which, because of the high energy needed, will most of the times be a valence quark ($90\%$ of $ug$ initial partons for $M=1.0$ TeV with $l^+l^+$ final state). The difference in the $(++)$ and $(--)$ cross--section is shown in figure \ref{figChargeAsymmetry} for $M_{T,B}=0.5-1.5$~TeV and $\lambda_{T,B}=2,3,4$. The difference in the cross-sections, rather than the asymmetry, could be a better observable in our case because the sign background contribution, as well as the one from pair production, disappears. In the asymmetry it enters in the denominator. For known $M_T$ and $M_B$, the charge asymmetry will put a constraint on the $\lambda_{T,B}$ couplings, we will come back on this later.

\begin{figure}[!tb]
	\centering
	\begin{tabular}{cc}
		\parbox[b]{6.5cm}{\hspace{-0.5cm}\rotatebox{90}{\scalebox{0.4}{\includegraphics{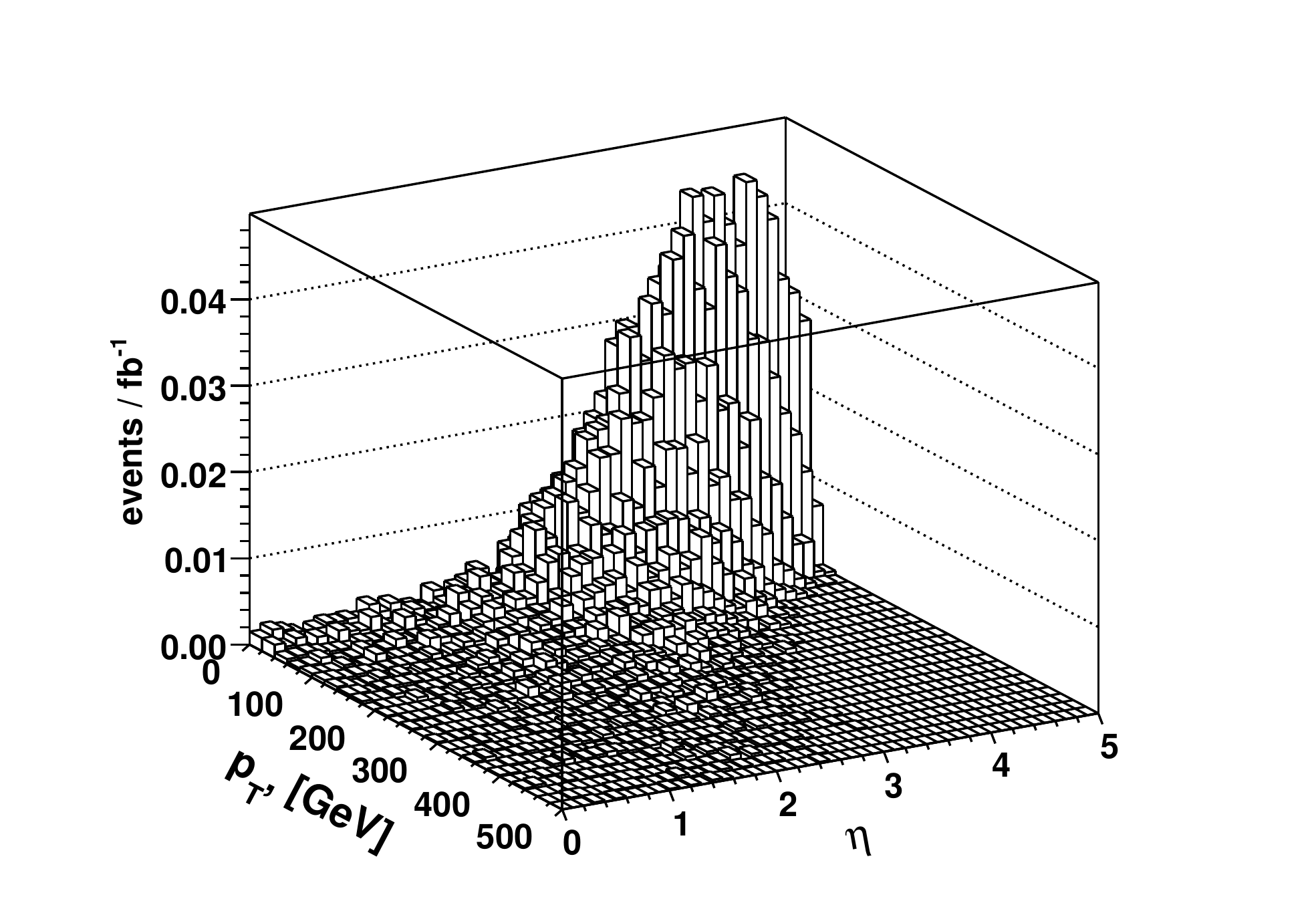}}}} &
		\parbox[b]{5.5cm}{\rotatebox{90}{\scalebox{0.3}{\includegraphics{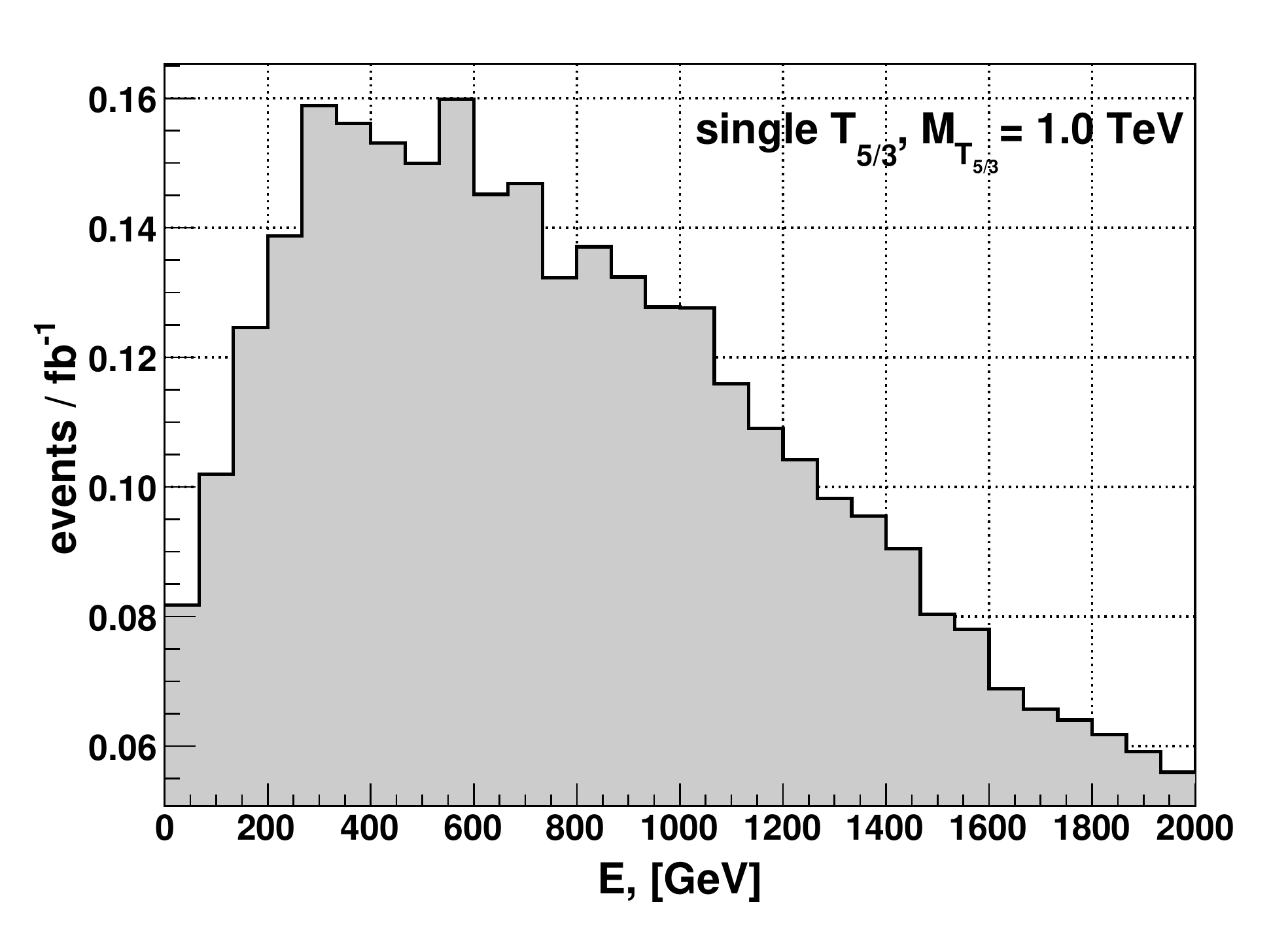}}}} \\
		\parbox[t]{6.5cm}{\hspace{-0.5cm}\rotatebox{90}{\scalebox{0.4}{\includegraphics{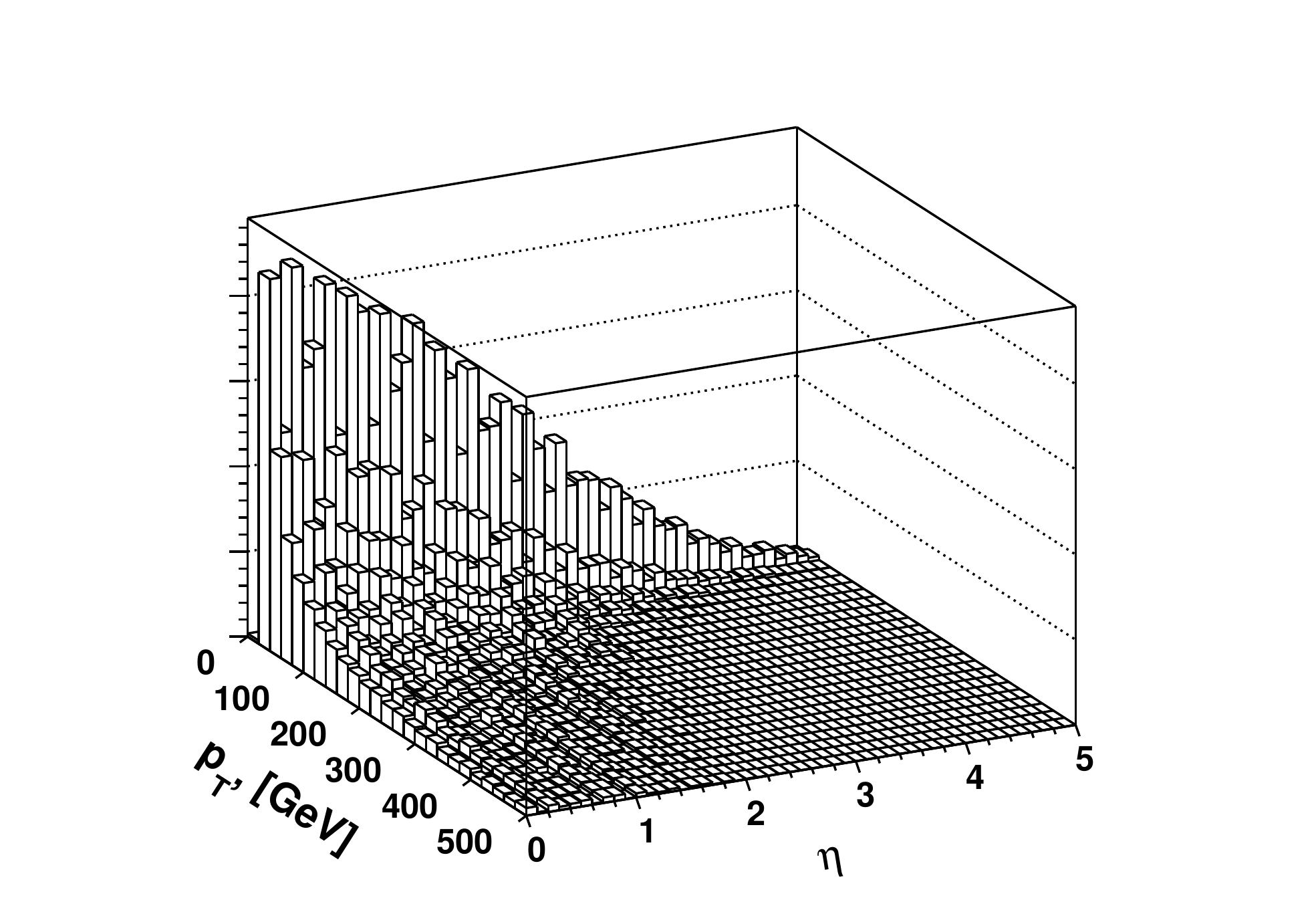}}}} &
		\parbox[t]{5.5cm}{\rotatebox{90}{\scalebox{0.3}{\includegraphics{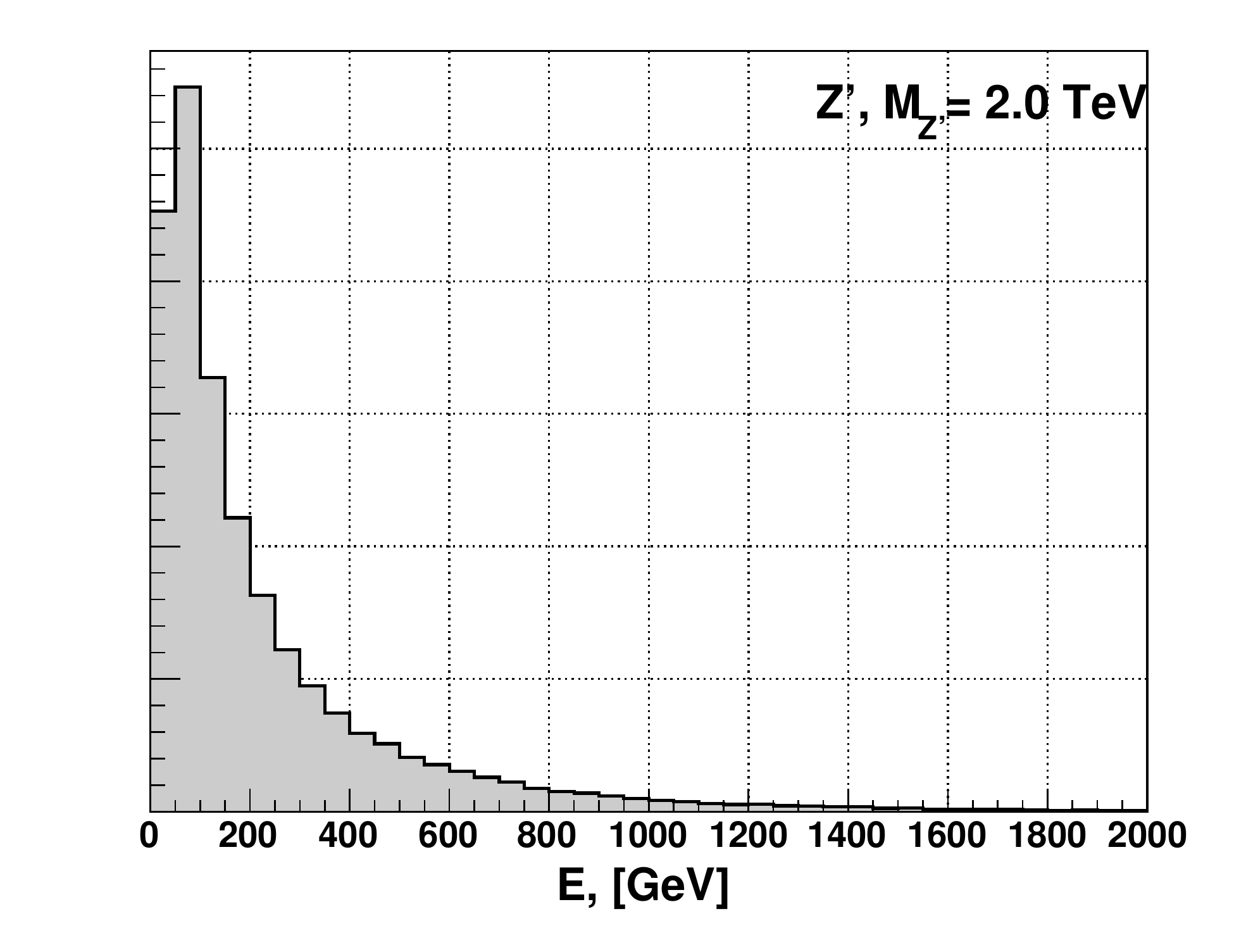}}}} \\
	\end{tabular}
	\caption{Distribution in the $\PT - \eta$ (left) plane and in energy (right) of the signal forward jet (top) and of the ISR jets (bottom).
		The signal is for single $T_{5/3}$ ($M_T=1.0$~TeV, $\lambda_T=3$) production, while ISR was simulated with a 2~TeV $Z'$.}
	\label{figFwdJet}
\end{figure}

The presence of a forward energetic jet is, as we have discussed in section~3, another evidence of the single production. Initial state radiation (ISR) in the pair production signal or in the background, when constrained by our cuts to kinematical regions of high hard scale, produces forward jets which are similar to the one we want to observe. The ISR jets are the main background to the forward jet identification and it is therefore important to take this effect into account.  Figure~\ref{figFwdJet} shows the typical ISR distribution for a hard scale of $2$~TeV in the $\PT$--$\eta$ plane and in the jet energy. Compared with the signal (see again figure~\ref{figFwdJet}), the ISR jets distribution is peaked around softer and more central emissions even for very hard ($2$~TeV) processes. For technical reasons, we have obtained the ISR radiation distribution in figure~\ref{figFwdJet} by simulating the Drell--Yan production of  a $Z'$ boson of $2$~TeV mass. This allowed us to obtain quickly large samples of high hard scale events and therefore a readable 2d distribution, while the resulting shapes are independent of this technicality.
Figure~\ref{figFwdJet} suggests the following criteria to identify the forward jet. Firstly, it has to be energetic, $E>300$ GeV, but this will also be often true for the other hard jets of our signal and background. Among these candidates, we therefore look for low $\PT$ and forward jets by imposing $\PT<150$ GeV and $\eta > 2.5$. In case of multiple candidates, we take the most energetic one. The efficiency on the single production at $1.0$ TeV is $65\%$, while the fake forward jet probability from the backgrounds and pair production is around $20\%$. Most of this fakes come, as we have explained, from the ISR effects which are of course included in our simulation of the background and of the pair production.

\subsection{Identification of the top partners}

The evidences for the top partners discussed up to now are indirect,  we will now try to identify the particles responsible for the excess, which would provide a  direct proof of their existence. In what follows we will propose a method to detect the presence of the $T_{5/3}$ and/or of the $B$ particles and to extract their mass. If one of them is significantly lighter than the other, or more strongly coupled, it will appear as the only particle present. Else, our signal will contain a mixture of the two charges whose relative importance depends on the masses and on the couplings.

\paragraph{Charge:}

We would like first of all to establish whether our signal can originate from a heavy fermion with charge $-1/3$ (the $B$), $+5/3$ (the $T_{5/3}$) or from a combined contribution of both. We will make use of the fact that (see figure~\ref{figProdDiagram}) both leptons come from the partner decay in the case of the $\T$  while the second lepton comes from the other top (or $B$) leg in the case of the $B$. This will be used in different ways, depending on the mass.
\begin{figure}[!p]
	\begin{center}
	\begin{tabular}{cc}
		\scalebox{0.4}{\rotatebox{90}{\includegraphics{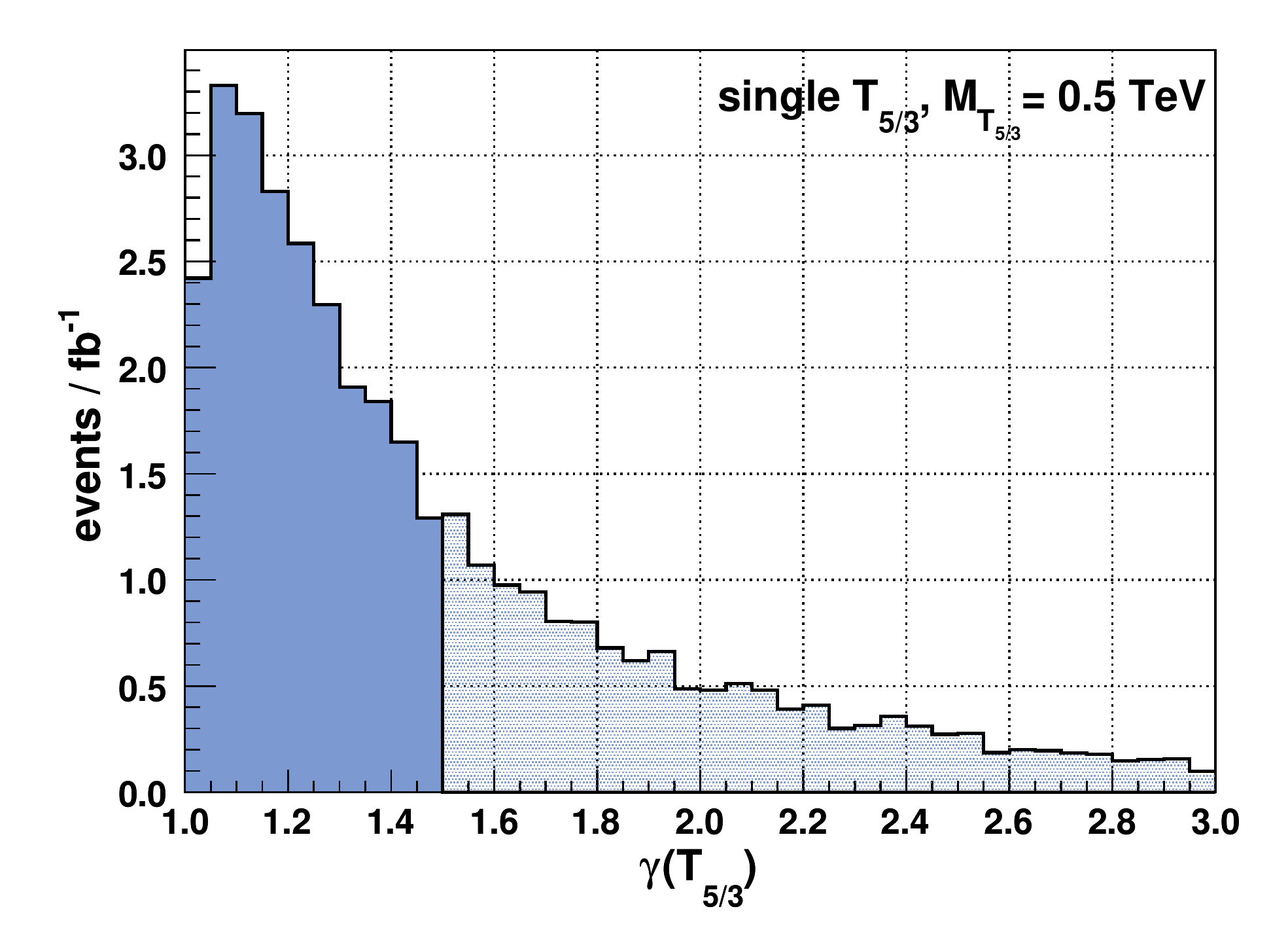}}} & 
		\scalebox{0.42}{\rotatebox{90}{\includegraphics{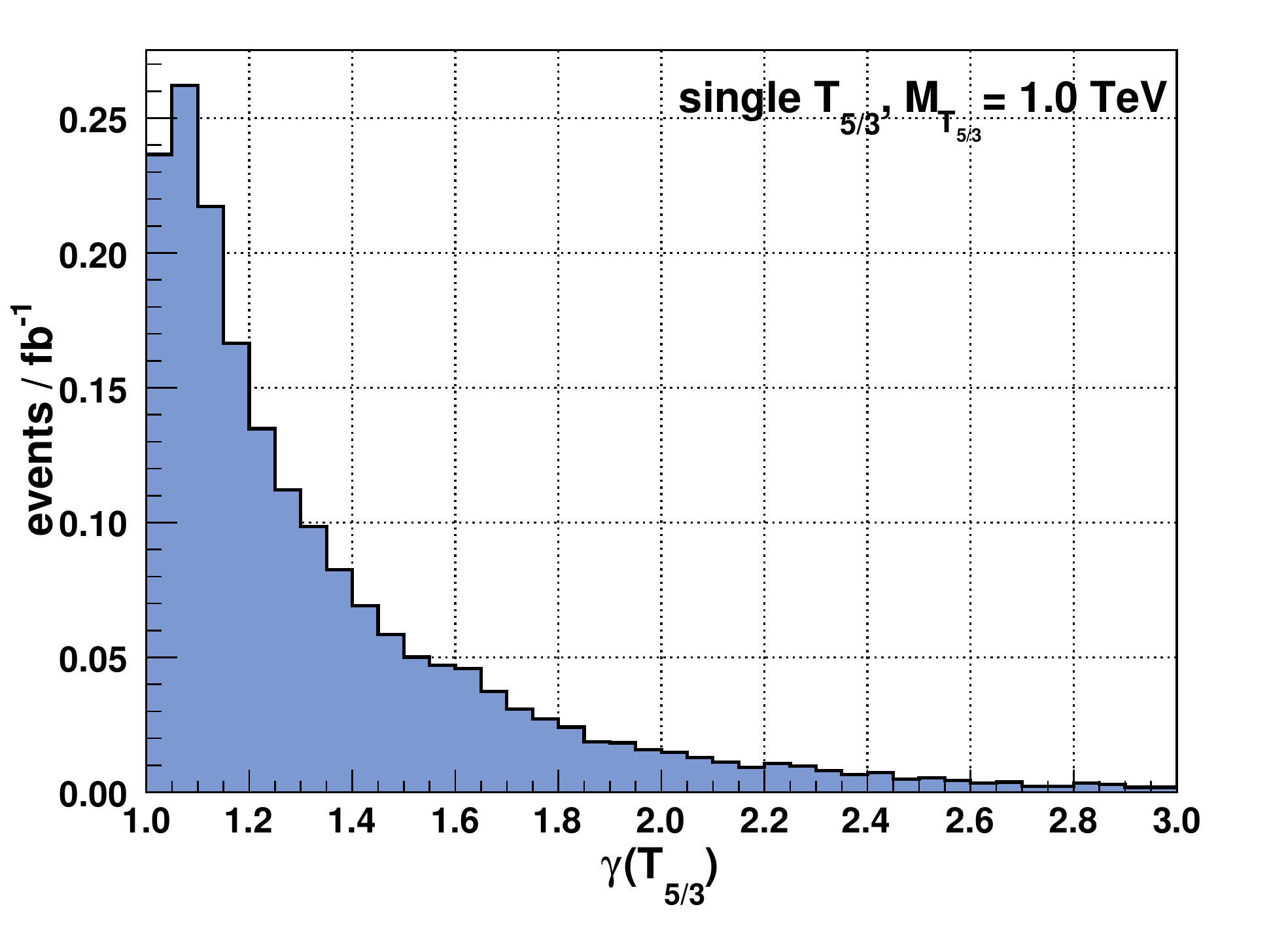}}}
	\end{tabular}
	\caption{The $\gamma$ factor for $\T$ in the case $\MT$ = 0.5 TeV or 1.0 TeV. The sample is divided into two bins, shaded for $\gamma >1.5$ and plain 	for $\gamma < 1.5$.}
		\label{figBoost}
   \begin{tabular}{cc}
		\scalebox{0.4}{\rotatebox{90}{\includegraphics{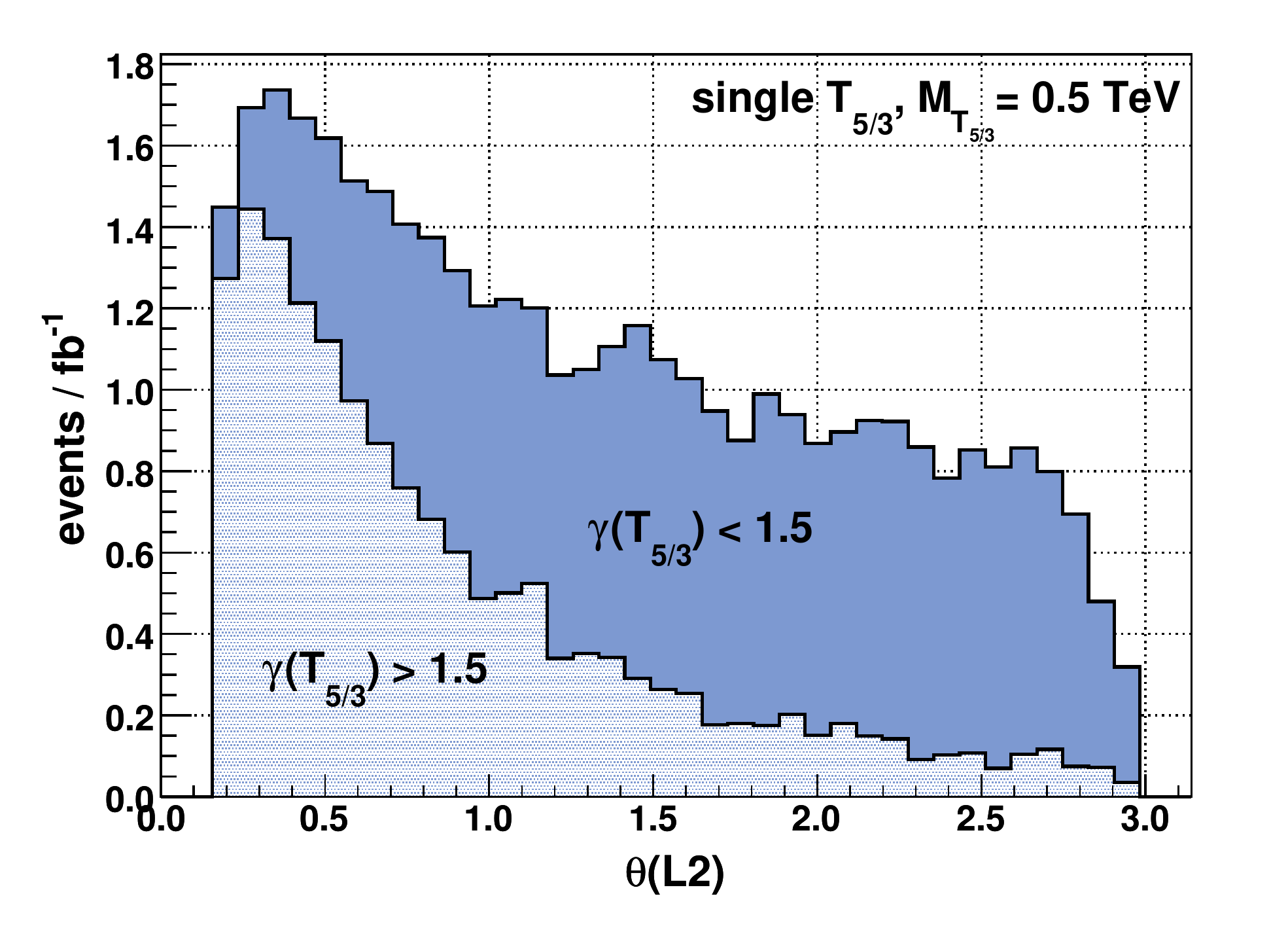}}} & 
		\scalebox{0.42}{\rotatebox{90}{\includegraphics{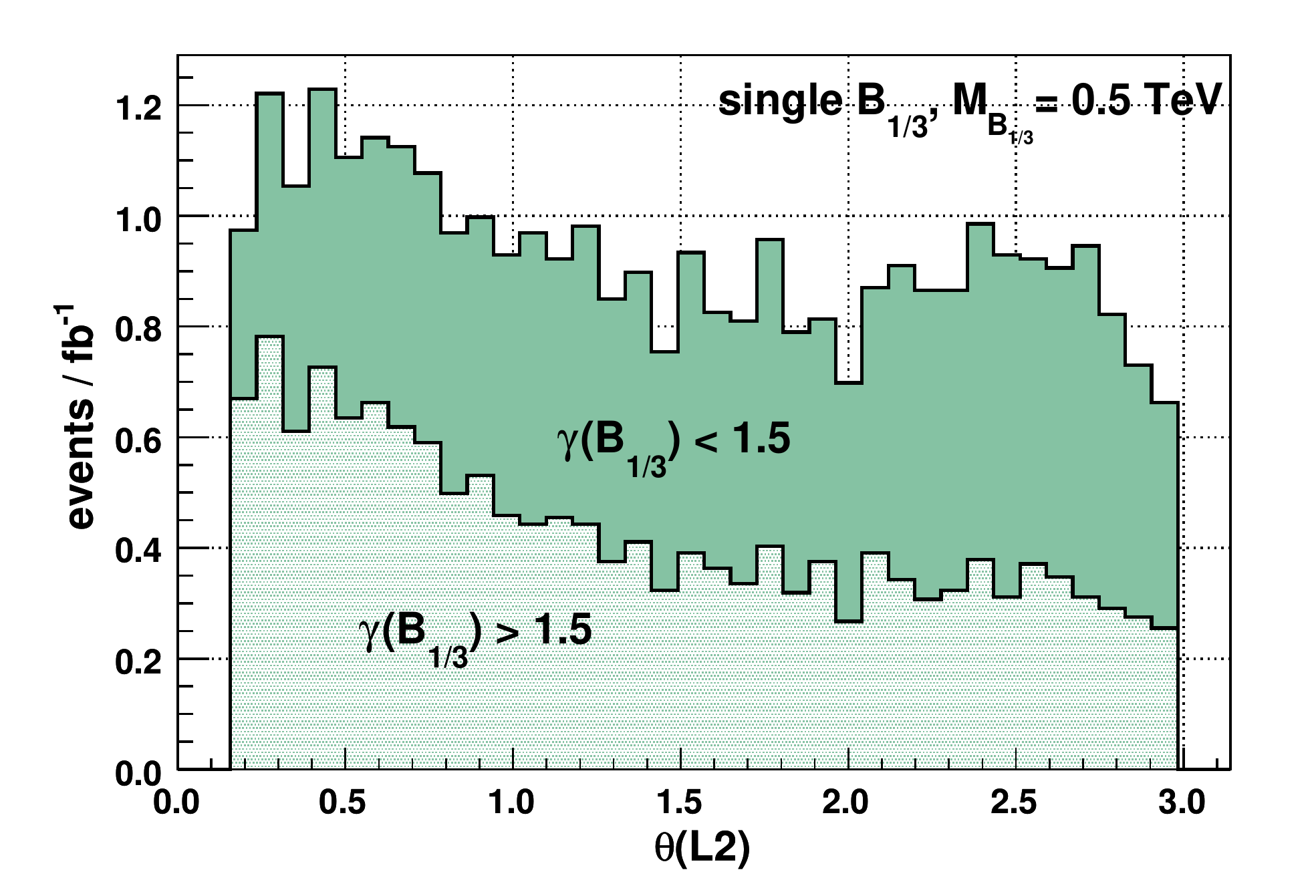}}}
	\end{tabular}
	\caption{The angle with the beam of the second lepton ($\theta(L2)$) for $\T$ and $B$, $M_{T,B}=0.5$ TeV. 
		The system is always oriented to have $\theta(L1)<\pi/2$.
		We see that boosted $\T$ leads to aligned leptons in the forward region, while if it is at rest the leptons tend to be opposite.
		The distribution is more symmetric in the case of the $B$.}
	\label{figThetaL2}
	\begin{tabular}{cc}
		\scalebox{0.4}{\rotatebox{90}{\includegraphics{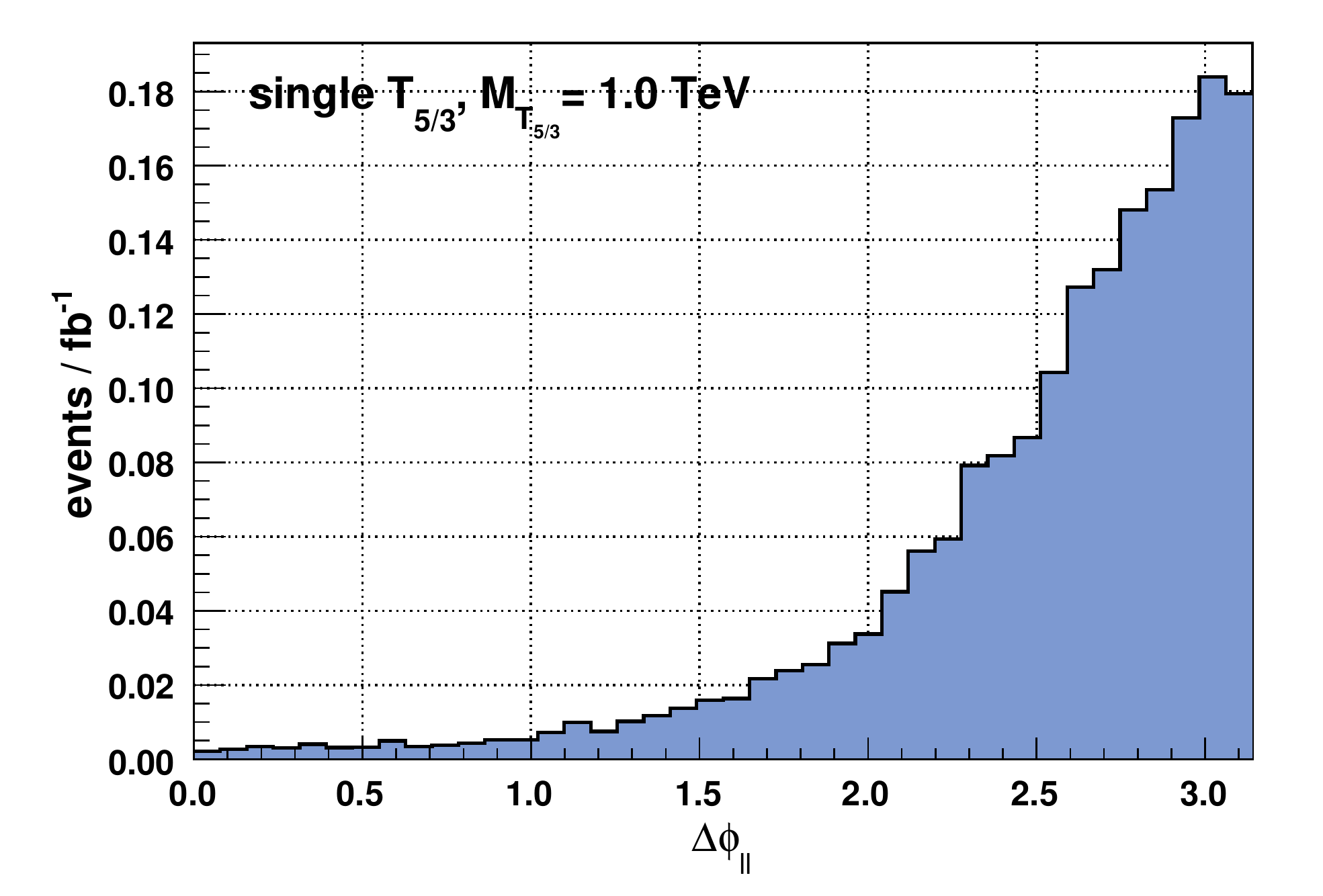}}} & 
		\scalebox{0.4}{\rotatebox{90}{\includegraphics{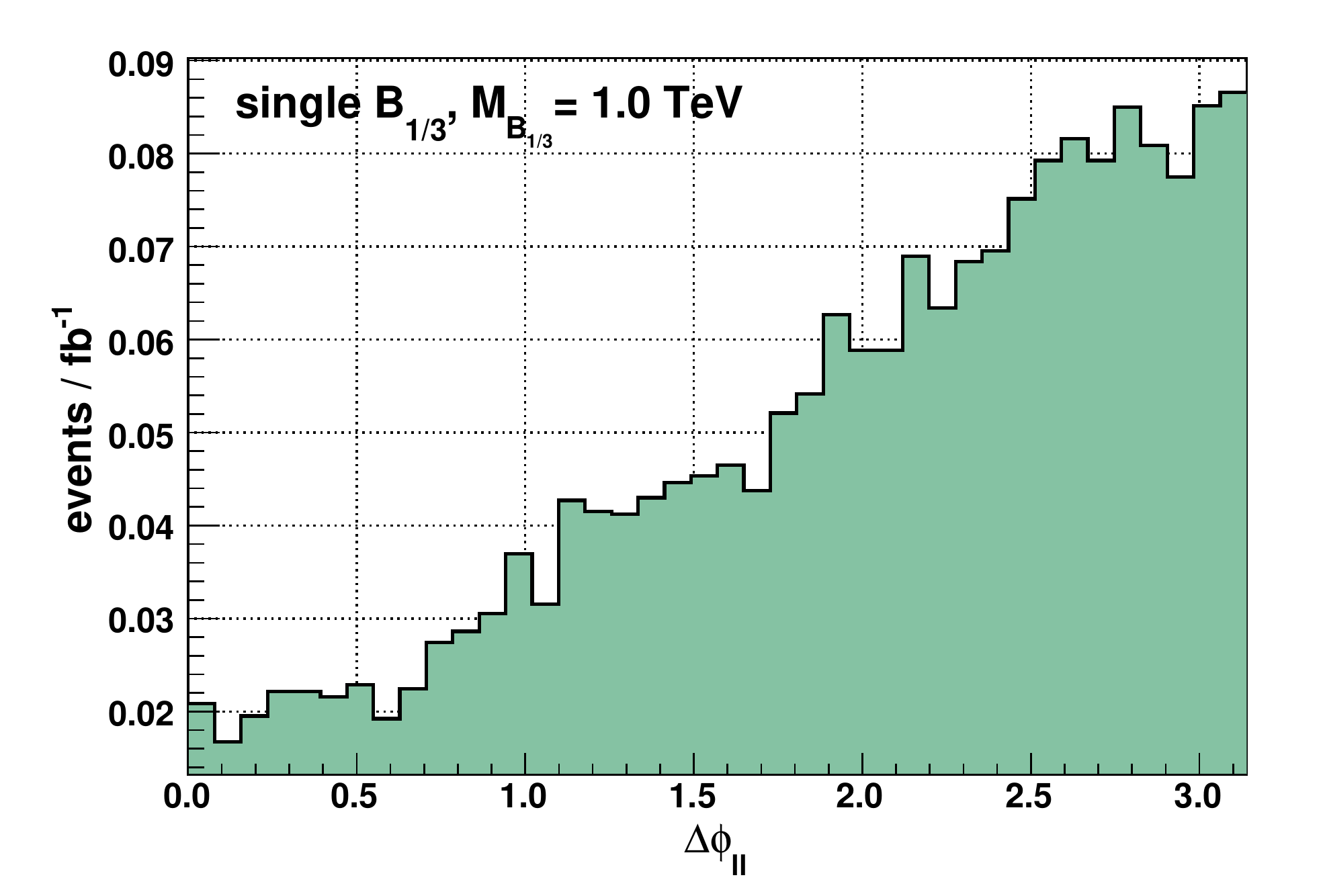}}}
	\end{tabular}
	\caption{$\Delta\varphi(LL)$ for $\T$ \& $\B$, $\MT=\MB=1.0$ TeV, medium cuts.
		The heavy particle tends to have low $\PT$, for the $\T$ this will imply that the leptons are back to back in the transverse plane as can be seen in the left plot. For the $\B$ one lepton comes from its decay while the other comes from the opposite top leg, their azimuthal angles are therefore less correlated.}
	\label{figDPhi}
	\end{center}
\end{figure}
For low masses, $M_{T,B}= 0.5$~TeV, the partners will be boosted along the beam axis (see figure~\ref{figBoost}) and when this is the case their decay products tend to be aligned and in the forward region. This is inherited, in the case of the $\T$, by the two leptons, which therefore are preferentially emitted in the same hemisphere. The opposite happens when the $\T$ is at rest since its decay products tend to be back to back in this case. This makes, as shown in figure~\ref{figThetaL2}, that the angle $\theta(L_2)$ of the second (in $\PT$) lepton with the beam, having oriented the system in such a way that $\theta(L_1)<\pi/2$, peaks at small values when the $\T$ is boosted and at large values when it is at rest. 
The $\T$ being preferentially boosted for $0.5$~TeV, the total distribution shows a preference for aligned leptons and, since this feature is not present in the case of the $B$, in can be used to distinguish the two. We therefore define the asymmetry between the events in which both leptons are in the same or opposite hemispheres ($\theta\lessgtr \pi/2$), $A_{ll}^\textrm{hemisphere} = (\#\textrm{same} - \#\textrm{opposite}) / (\#\textrm{all})$. The predicted asymmetry for $B$, $\T$ and for a combination of the two are shown in table \ref{tabRhemisphere}, we observe significantly different results in the various cases. 
\begin{table}[!tb]
	\begin{center}
	\begin{tabular}{|c|c|c|} \hline
	$A_{ll}^\textrm{hemisphere}$	& $\B$		& $\T$			\\ \hline
	$\T$ or $\B$			& 0.05		& 0.23			\\ \hline
	$\T$ and $\B$			& \multicolumn{2}{c|}{0.14}		\\ \hline
	\end{tabular}
	\end{center}
	\caption{Table giving the asymmetry between leptons in same or opposite hemisphere, $A_{ll}^\textrm{hemisphere}$, for the signal (single \& pair production) at $\MT=\MB=0.5$ TeV with soft cuts applied. The background contribution is negligible for low masses, therefore it has been ignored.
}
	\label{tabRhemisphere}
\end{table}

For higher masses, $M_{T,B} \gtrsim 1.0$ TeV, the partners will hardly be boosted and the hemisphere asymmetry will be washed out. To distinguish the $\T$ from the $B$ we can use, in this case, the fact that the partners usually have very small transverse boost as a result of the single production mechanism. Moreover, since they are heavy, the top and the $W$ from their decay will be significantly boosted so that they will transmit their direction to their decay product.
In the case of $\T$, therefore, the leptons will preferentially be back to back in the transverse plane, as shown in figure \ref{figDPhi}. Notice that the leptons tend to be separated in the transverse plane also for the $B$, due to the fact that the $B$ and the leptonic top in the single production diagram preferentially go in opposite transverse directions. This feature is however less sharp for the $B$ and can be used for the identification of the partners. The ratio between opposite lepton pairs and the aligned ones, $R_{ll}^{\Delta\varphi}=(\#\Delta\varphi_{ll}>2.5) / (\#\Delta\varphi_{ll}<1.0)$, is reported in table \ref{tabRDphi} and seems a promising observable to distinguish among the partners.

\begin{table}[!tb]
	\begin{center}
	\begin{tabular}{|c|c|c|} \hline
	$R_{ll}^{\Delta\varphi}$		& $\B$		& $\T$			\\ \hline
	background			& \multicolumn{2}{c|}{2.95}		\\ \hline
	($\T$ or $\B$) + background	& 1.98		& 6.00			\\ \hline
	$\T$ + $\B$ + background	& \multicolumn{2}{c|}{3.33}		\\ \hline
	\end{tabular}
	\end{center}
	\caption{Table giving $R_{ll}^{\Delta\varphi} = (\# \Delta\varphi_{ll} > 2.5) / (\# \Delta\varphi_{ll} < 1.0)$ for $\MT=\MB=1.0$ TeV, with medium cuts applied.
	Significantly different results are obtained in the various configurations with only the $\T$, only the $B$ or a superimposition of the two.
	}
	\label{tabRDphi}
\end{table}

\paragraph{Mass:}

Let us now discuss some strategies to measure the top partners mass. The first method that could be employed is based on the ``$\mtt$--assisted'' missing momenta reconstruction proposed in \cite{Cho:2008tj-bulk}, which is suited for our case since the only sources of $\ET$ are the two neutrinos from the $W$ decays. This is based on the possibility of reconstructing the neutrino's momenta, and therefore also the $W's$, in the events which lie close to the $m_W$ threshold of the $\mtt$ distribution of figure~(\ref{figMt2W}), and once these are known the top partners could be reconstructed both in the single and pair production cases (see figure~\ref{figProdDiagram}) by reconstructing their decay products.
In all cases we need to identify a $b$ quark not belonging to a hadronic top, and for the $\B$ we also need 
to establish which of the reconstructed $W$ originates from a semi-leptonic top. This could be achieved by 
requiring the $b+W$ invariant mass to be close to $m_t$.
We will not study this interesting possibility in detail, but we will rather describe other methods to measure the top partner mass based on more standard observables.

For the $T_{5/3}$, two strategies can be employed. If it is pair produced, the $T_{5/3}$ which produces the leptons is accompanied  by a second $T_{5/3}$ with hadronic decay (see figure~\ref{figProdDiagram}). The latter can be reconstructed and its mass measured as proposed in \cite{Contino:2008hi}. A second method, not based on pair production and therefore suited also for high masses where the pair production cross--section is low, is based on the usual transverse mass $\mt$ \cite{Amsler:2008zzb}, extended to more than two particles
\begin{equation}
	\mt = \sqrt{\Big(  \sum_{i} \PT\mbox{}_i\Big)^2 - \Big( \sum_{i} \vec\PT\mbox{}_i \Big)^2 }\,,
\end{equation}
where the sum will run over $i=\ET, L_1, L_2, b$, the $b$ being defined as a $b$-jet (identified with an efficiency of $\epsilon_b\approx0.5$) not belonging to the hadronic top. Our $\mt$ is an underestimate of the true $\T$ transverse mass, since the sum of the neutrinos transverse energies has been replaced with $\ET=|\slashed{\vec{E}}_T|=|\vec{p}_T(\nu_1)+\vec{p}_T(\nu_2)|\leq |\vec{p}_T(\nu_1)|+|\vec{p}_T(\nu_2)|$. The central relation $\mt < \MT$ is therefore satisfied and we can use the end--point of the $\mt$ distribution to measure the $T_{5/3}$ mass. The main background which affects this distribution is the other partner, we can however obtain a cleaner sample of $T_{5/3}$ events by the cut $\Delta\varphi_{ll}>2.5$ (see figure \ref{figDPhi}). We will describe a concrete example of this method in the following subsection.

In the case of the $B$, and for single production, we can use the $\mtt$ variable introduced above (eq.~(\ref{eqMt2})), but in a non-standard way since we will apply it to a system of non--degenerate particles, the $B$ and the leptonic top (see figure~\ref{figProdDiagram}). To assign each lepton to the right leg we will use that the one coming from the decay of the $B$ will often be the hardest one (for $\MB=1.0$ TeV this is true the $83\%$ of the cases). The hardest lepton will therefore be combined with the hadronic top, reconstructed as discussed above, and constitutes the first visible decay product . Therefore, in the notation of eq.~(\ref{eqMt2}), we have $q_1 = p(L1) + p(\textrm{hadronic top})$. For the second particle we require a $b$ jet (paying again an efficiency $\epsilon_b \approx 0.5$) and we combine it with the second lepton: $q_2 = p(L2) + p(b)$. The end--point of the $\mtt$ distribution is the heaviest of the two decaying particles, namely $M_B$. The biggest background for the $M_B$ determination is the $\T$ contribution, which can be lowered by selecting events with $\Delta\varphi_{ll}<2.5$, see figure \ref{figDPhi}.

\begin{figure}[!t]
	\centering
	\begin{tabular}{cc}
		\multicolumn{2}{c}{$\MT=1.0$ TeV, $\MB=0.8$ TeV} \\
		\scalebox{0.3}{\rotatebox{90}{\includegraphics{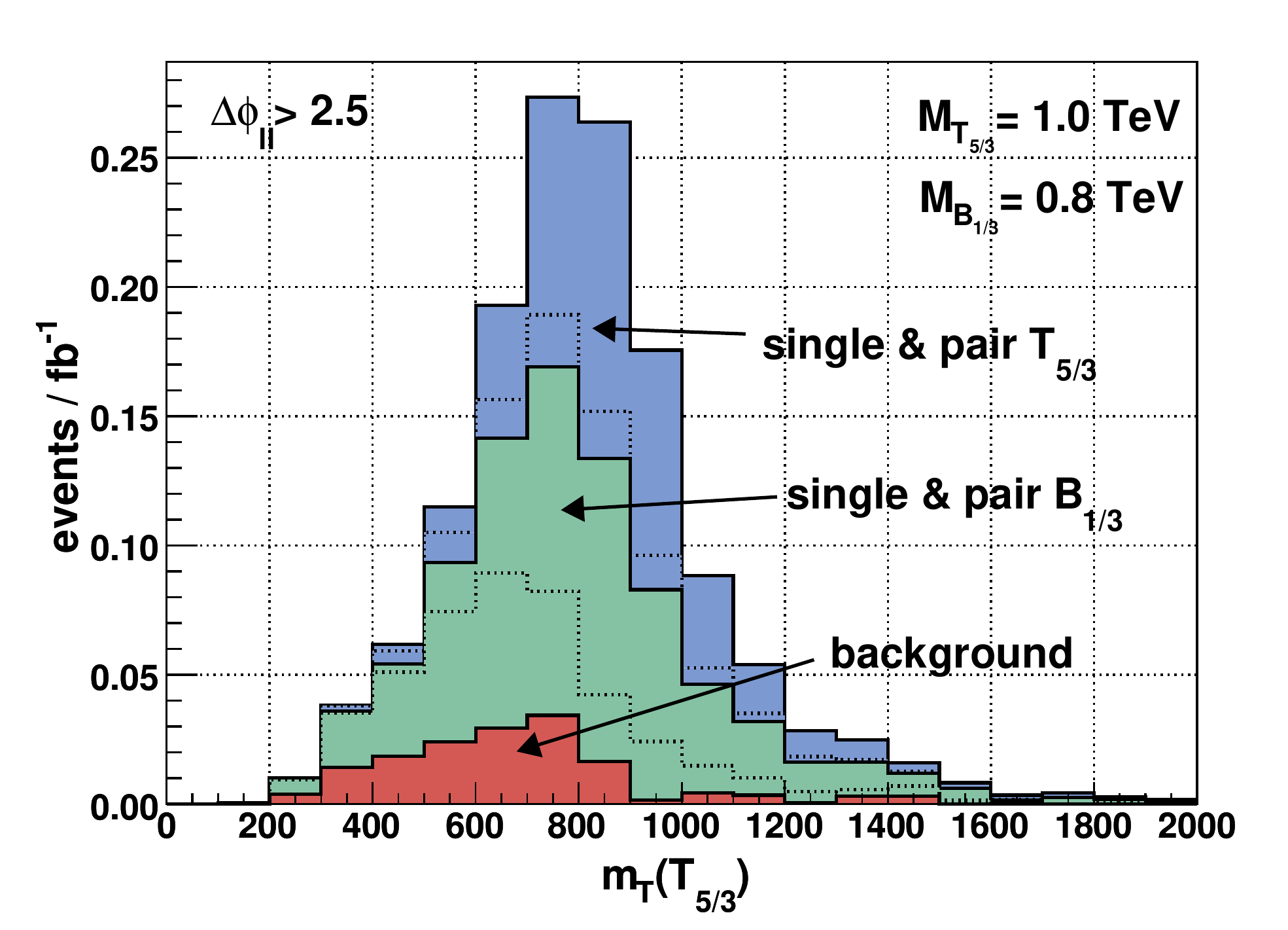}}} &
		\scalebox{0.3}{\rotatebox{90}{\includegraphics{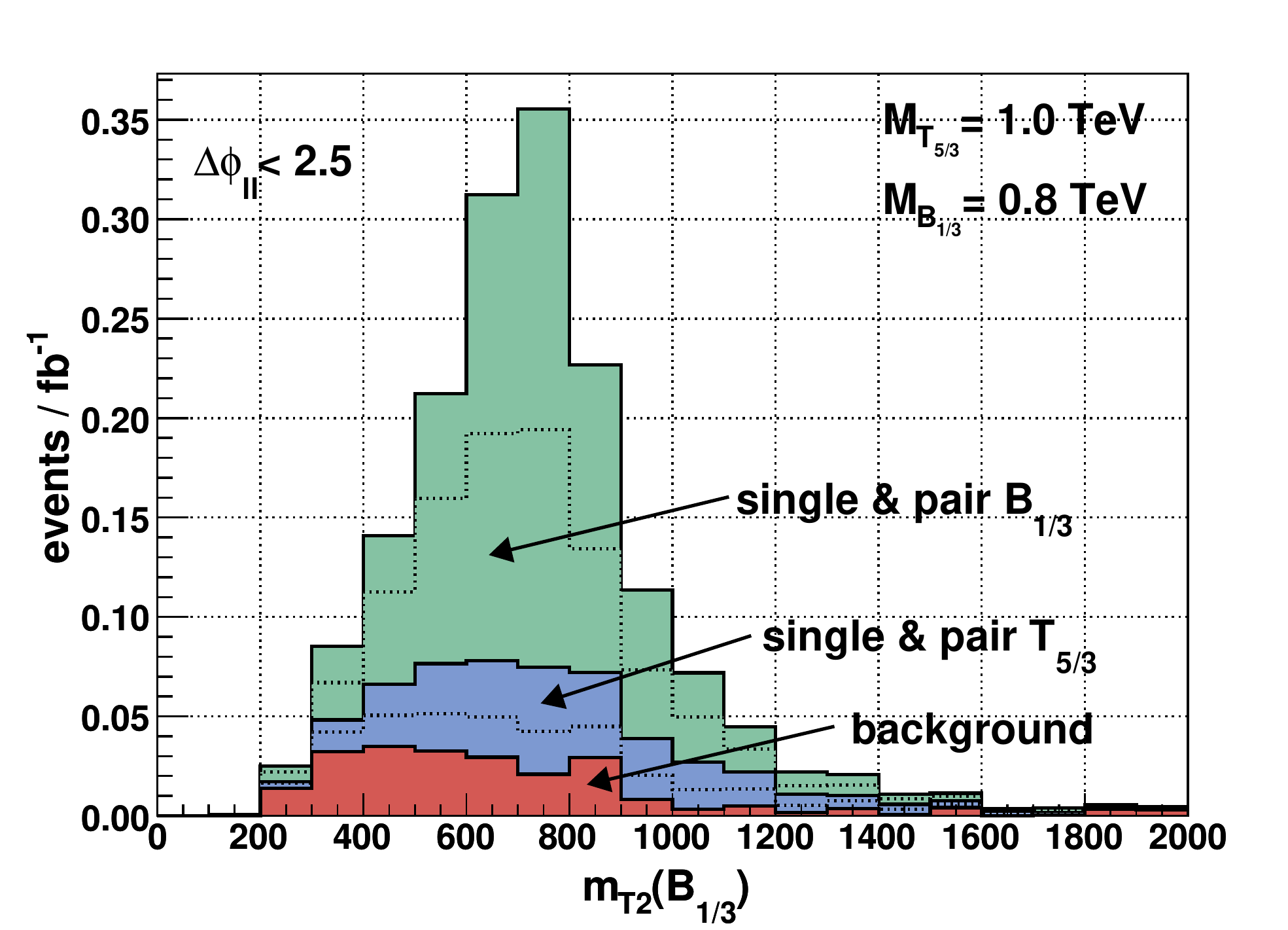}}} \\
		(a) $\mt(\T)$ for $\Delta\varphi_{ll} > 2.5$ & (b) $\mtt(\B)$ for $\Delta\varphi_{ll} < 2.5$\\ \\ \\
		\multicolumn{2}{c}{$\MT=1.0$ TeV, $\MB=1.2$ TeV} \\
		\scalebox{0.3}{\rotatebox{90}{\includegraphics{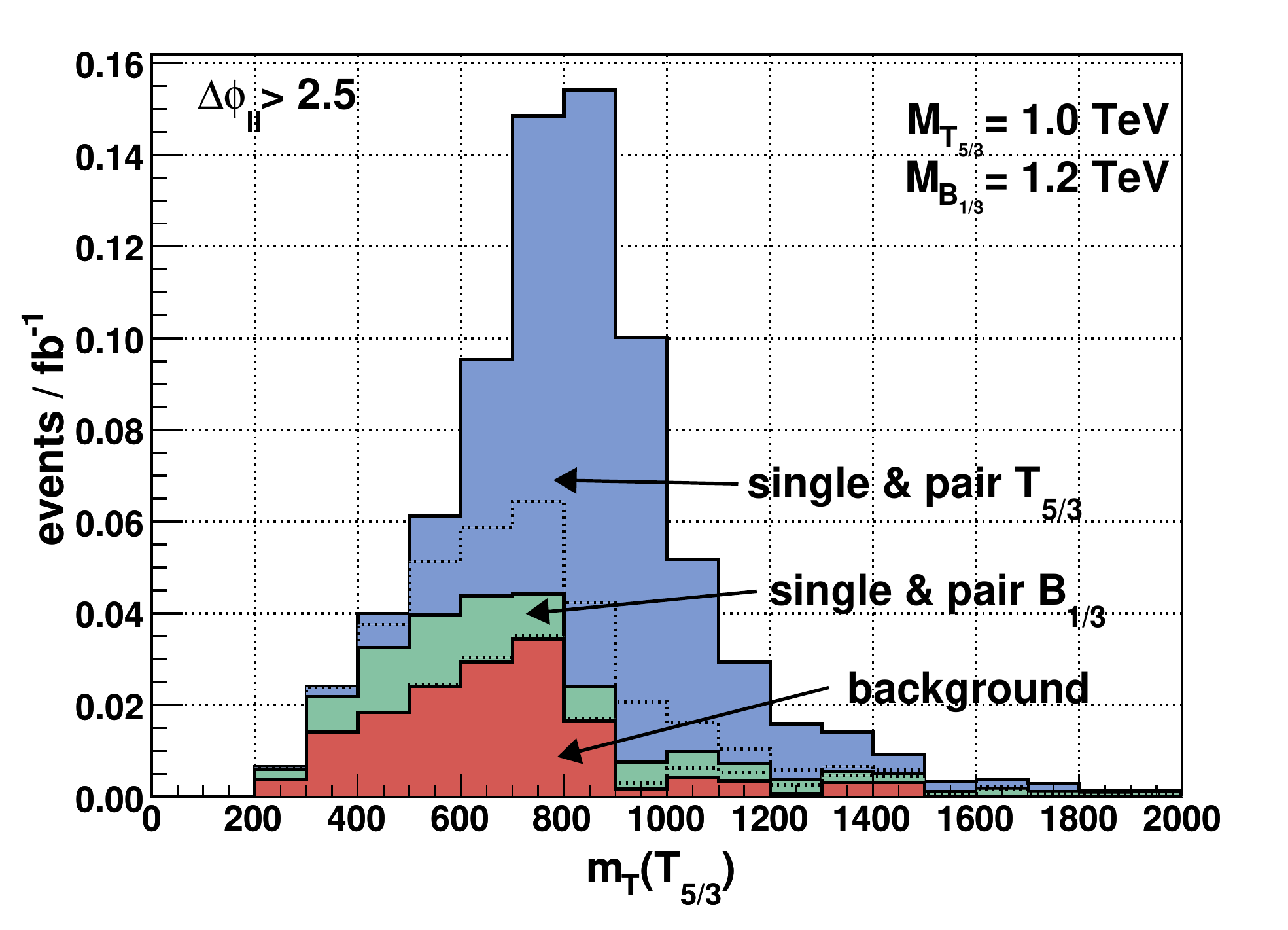}}} &
		\scalebox{0.3}{\rotatebox{90}{\includegraphics{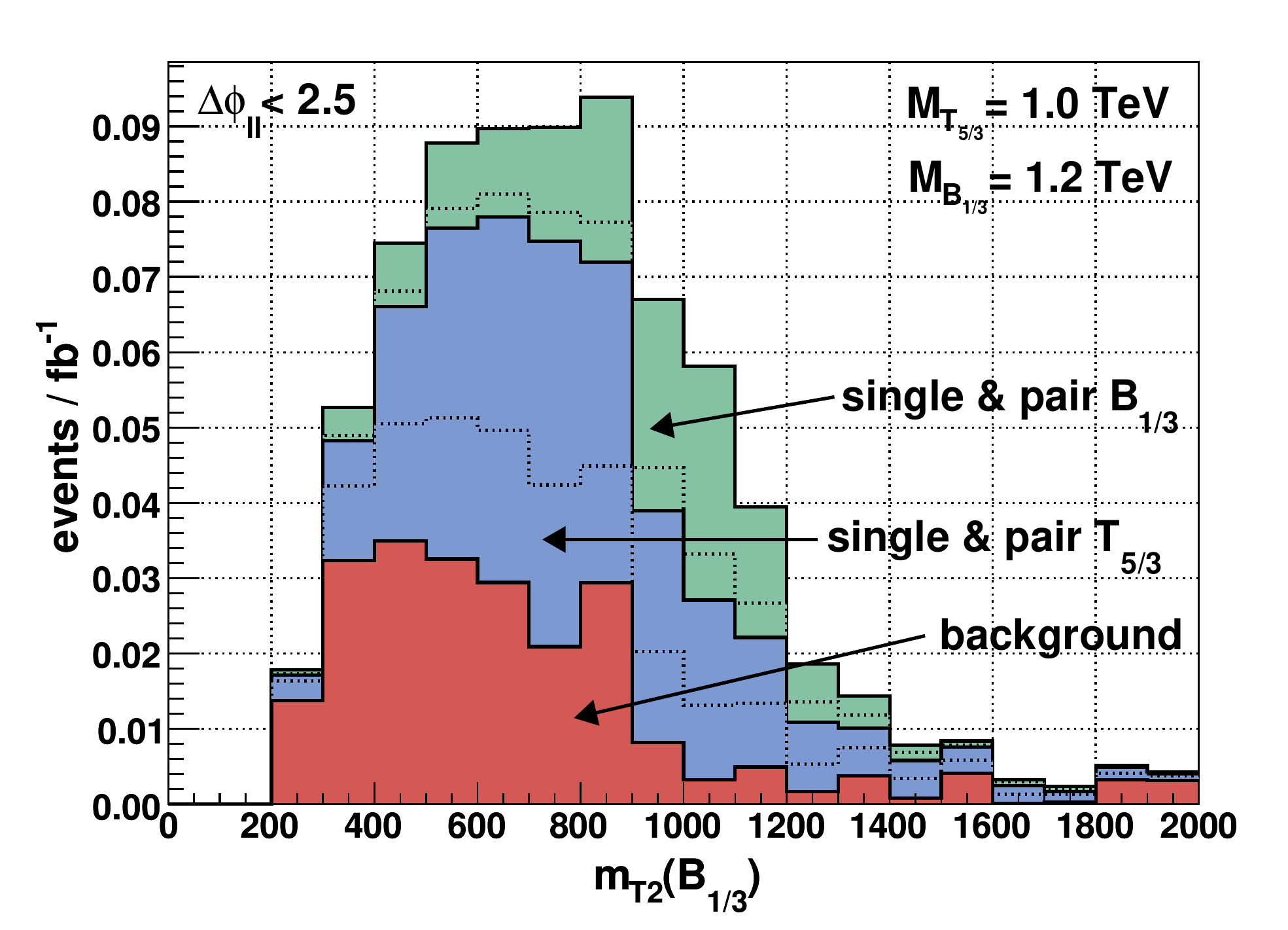}}} \\
		(c) $\mt(\T)$ for $\Delta\varphi_{ll} > 2.5$ & (d) $\mtt(\B)$ for $\Delta\varphi_{ll} < 2.5$\\
	\end{tabular}
	\caption{$\mt(\T)$ and $\mtt(\B)$ distributions constructed as defined in the text, with respectively $\Delta\varphi_{ll}\gtrless 2.5$.
		These plots show the results only for the $l^+l^+$ channel after medium cuts.
		$b$ tagging efficiency is taken into account.
		We notice how well the $\Delta\varphi_{ll}$ selection works.
		For the $\mt(\T)$ we look for a peak followed by a threshold at $\MT$, while for $\mtt(\B)$ we observe for a sharp decrease at $\MB$.
		It seems that all masses could be extracted.}
	\label{figMassDistributions}
\end{figure}

\paragraph{Example:} The methods described in this section should allow to easily distinguish the situations in which only the $B$ or only the $\T$ are present, and to measure their mass. In order to test our ability to disentangle the $B$ and $\T$ effects in more subtle situations, we consider now the case in which \mbox{$M_T=1.0$}~TeV and $M_B=0.8(1.2)$~TeV and $\lambda_{T,B}=3$. We will see that the identification of the two particles is more difficult but still possible. 
The total cross-section after cuts for the signal is $12.2$~fb ($5.0$~fb), while for the background we have $2.5$~fb, which leads to $L_\textrm{disc}=490$~pb$^{-1}$ ($2.4$~fb$^{-1}$).
The different mass distributions are shown in figure \ref{figMassDistributions}, they indicate that it should be possible to extract the mass of both exotic quarks even in these intermediate situations using the different transverse semi-leptonic mass distributions defined above.

\begin{figure}[!h]
	\centering
	\scalebox{0.5}{\rotatebox{90}{\includegraphics{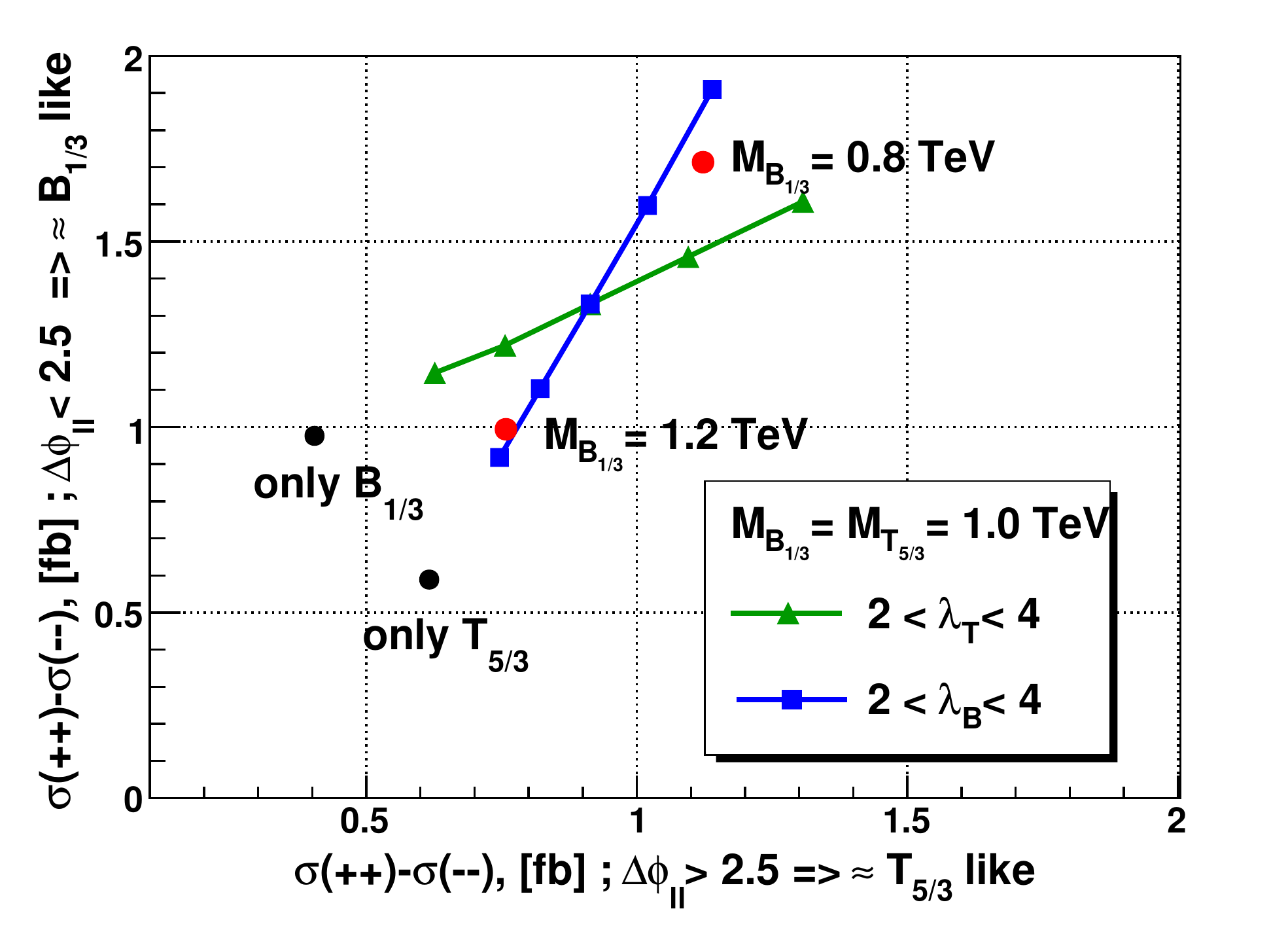}}}
	\caption{Charge asymmetry for events with $\Delta\varphi_{ll}>2.5$ (rather $\T$ like) and $\Delta\varphi_{ll}<2.5$ (rather $\B$ like), for $\MT=\MB=1.0$ TeV and varying $\lambda_{T}$, $\lambda_{B}$.
	The bigger $\lambda$, the larger asymmetry since single production is enhanced.
	The two points with only $\T$ or $\B$ with $\lambda_{B,T}=3$, and the two points with different values of $\MB$, described in the example above, are also shown.
	We see that these two variables have distinct dependence in $\lambda_{T}$, $\lambda_{B}$ and could therefore in principle be inverted to measure them independently.
	}
	\label{figAsymCross}
\end{figure}

\paragraph{Coupling constants: }
If only one partner is present, its nature could be established and its mass measured in the way we have discussed. Once this is done, its coupling constant could be extracted from the charge asymmetry, see figure \ref{figChargeAsymmetry}. When both partners are present, we will still be able to measure their mass, but even when the latter is known one more observable will be needed, on top of the charge asymmetry, to measure their couplings.  One possible strategy would be to measure the charge asymmetry (or better, the cross--section difference)  for the two sets of events with $\Delta\varphi_{ll} \lessgtr 2.5$. Since we have already seen that the transverse angular separation is (for $M{T,B}\sim1.0$ TeV, at least) an efficient criterion for separating the two contributions, the two observables will show different sensitivities to $\lambda_T$ and to $\lambda_B$. This is shown in figure~ \ref{figAsymCross}.

\newpage

\section{Lower beam energy}

In our analysis we have assumed the LHC to work at its design center--of--mass energy of $\sqrt{S}=14$ TeV, and there is no reason to doubt that this energy will be reached at some stage of the LHC program. It is known, however, that the program will start with lower beam energies, so that it is worth discussing how our result will be modified in that case.
\begin{figure}[!htb]
	\begin{center}
	\scalebox{0.4}{\rotatebox{90}{\includegraphics{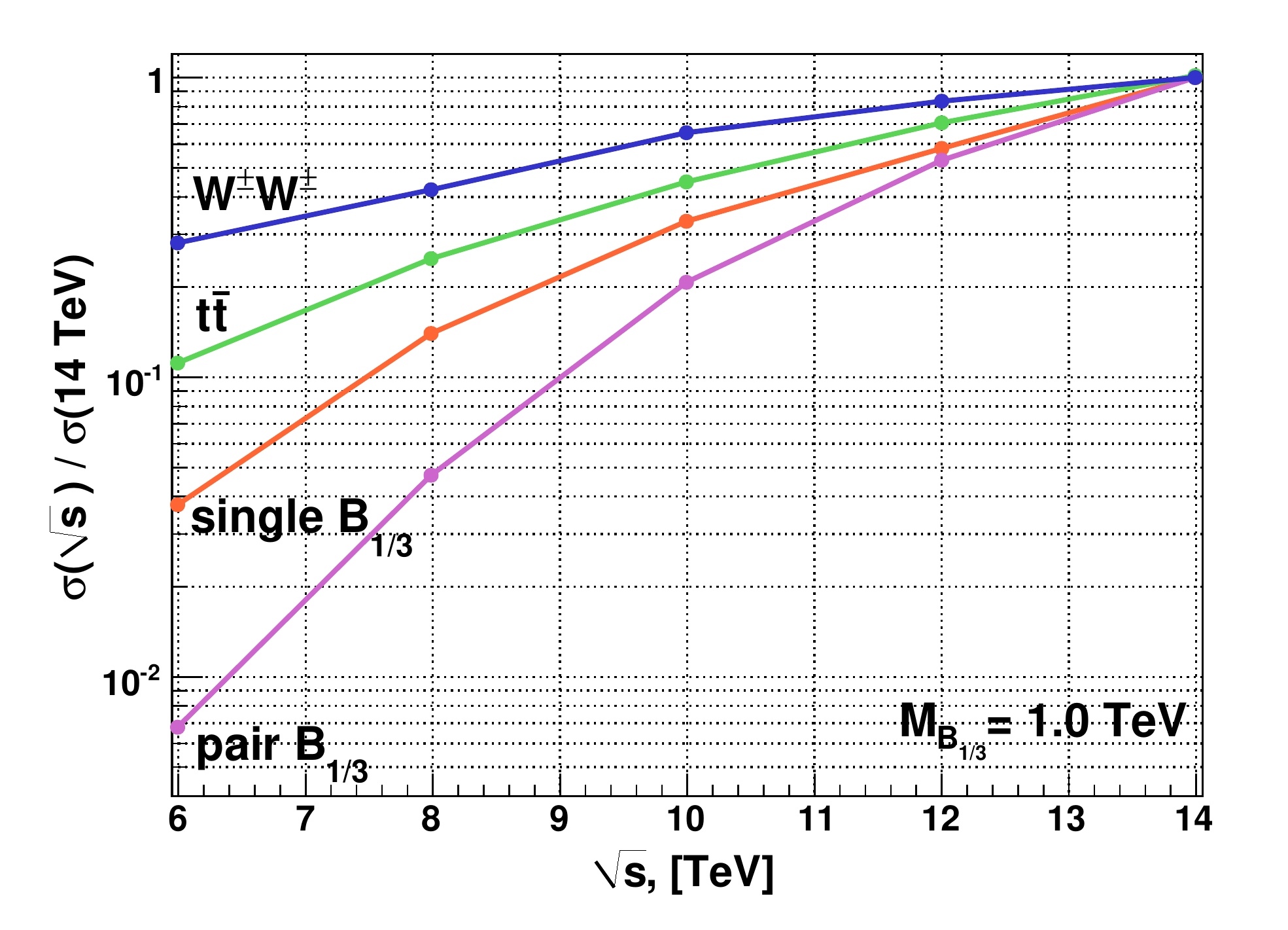}}}
	\end{center}
	\caption{Dependence on the beam energy of the cross section for single \& pair $\B$ ($\MB$=1.0 TeV), $t\bar t$ and $W^\pm W^\pm$.
		They are normalized to their 14 TeV value.
		We observe that the higher the intrinsic scale is, the more it is suppressed at low beam energy.}
	\label{figXsecBeam}
\end{figure}
Two mechanisms are at work when lowering the beam energy. First, all the cross-sections scale down, but those with a high intrinsic scale decrease \mbox{faster} as figure \ref{figXsecBeam} shows. For instance, while single $\B$($1.0$ TeV) goes down by a factor $\approx 4.0$, $t\bar t$ only decreases of a factor $\approx 2.3$ for $\sqrt{s} = 14 \to 10$~TeV. Second, since our cuts are only based on the hard transverse dynamics resulting from the decay of the partners, their efficiency  on the signal will only depend on the mass and not on the beam energy.\\
For the background, on the contrary, the efficiency of the cuts decreases with the beam energy since hard background events which could pass the cuts become less frequent. As an example, medium cuts on $tt$ have an efficiency of $2.6\cdot 10^{-2}$ at $14$~TeV and of $1.7\cdot10^{-2}$ at $10$~TeV.

It is therefore not worth modifying the cuts for lower energies. The discovery luminosities at $10$~TeV energy in the case $\MB=\MT$ and $\lambda_{T,B}=3$, presented in table \ref{tab10TeVLdisc}, are obtained with the same cuts discussed in section~3.
\begin{table}[!b]
\centering
\begin{tabular}{|c|r|r|r|r|r|}\hline
\multirow{2}{2cm}{ Mass, [TeV]} & \multicolumn{2}{c|}{$\sigma$, [fb]} & \multirow{2}{2.5cm}{$L_\textrm{discovery}$, [fb$^{-1}$]} & \multicolumn{2}{c|}{$\#$ events}	\\ \hhline{|~|--|~|--|}
		& signal		& background		&				& signal	& background		\\  \hline
	0.5	& 69.8\phantom{0}	& 10.9\phantom{00}	& 0.072				& 5		& 0			\\  \hline
	1.0	& 1.64			& 0.65\phantom{0}	& 5.5\phantom{00}		& 9		& 3			\\  \hline
	1.5	& 0.11			& 0.092			& 210.0\phantom{00}		& 22		& 19			\\  \hline
\end{tabular}
\caption{Cross-sections and discovery luminosity for $\MB=\MT=$ 0.5, 1.0 \& 1.5 TeV, with a beam energy at 10 TeV.
	The cuts used are the same as at 14 TeV, which leads to similar signal over background and hence lower significance.
	They are nevertheless still relevant.}
\label{tab10TeVLdisc}
\end{table}
The table shows that for $\MB=\MT=$ 0.5 or 1.0 TeV the discovery will be relatively easy, and that even for $\MB=\MT= 1.5$~TeV discovery will be possible if the entire programmed luminosity of $300$~fb$^{-1}$ will be collected at this energy.

Under the assumption that enough luminosity will be collected, the phenomenological study of the top partners and the measure of couplings and masses will still be possible, we have checked that the discussion of section~4 is still valid, for $10$~TeV energy, with minor numerical modifications.

\section{Conclusions}

We have studied the possibility of observing heavy partners of the top quark at the LHC, in the channel of same--sign dileptons. The top partners constitute a robust consequence of the partial compositeness hypothesis, which is realized in the compelling 5d (or 5d--inspired) models of strong--sector EWSB. We have seen that, under wide and motivated assumptions, the top partners are expected to couple strongly to the top quark through a vertex which also contains a longitudinal $W$ boson, and that this vertex is responsible for a sizable single production cross--section. The latter dominates, for high top partner mass, over the QCD--mediated pair production and therefore extends the discovery reach of the LHC. We have found that the discovery will be possible up to at least $1.5$~TeV in almost the entire expected 
parameter space.



Top partners below $1.5$~TeV are, as discussed in the Introduction, very likely to be present in both the Higgsless and in the composite Higgs scenarios, so that the one of same--sign dileptons is found to be a very promising channel in which these models could be tested at the LHC. For the composite higgs case, in which all other new states are expected to be heavier than the top partners which lie in the $[0.5,1.5]$~TeV range, the signal we have studied could constitute the most accessible experimental prediction. 

We have also discussed how, after the discovery of an excess, the presence of the top partners could be detected and their masses and couplings measured. Single production plays, also in this case, a major role since it allows to distinguish the partners from generic colored heavy fermions and to measure their couplings.

The above results have been established by performing a quite detailed simulation, using the \MGME~tools with showering performed with \Pythia~and 
the emission of extra partons taken into account by the MLM matching prescription. However, no genuine higher order corrections have been taken into account while these are expected to be sizable \cite{Bonciani:1998vc} both for the signal and for the background. Also, we have not included detector effects apart from charge misidentification and fake $\ET$. A more detailed analysis, for which we hope our results constitute a valid starting point, should clearly include a full detector simulation and the above--mentioned radiative corrections should be taken into account.

\section*{Acknowledgments}

We are indebted with R.~Contino, R.~Franceschini and especially with R.~Rattazzi for the many useful discussions and suggestions. We thank M.~Pierini for instructive conversations and especially for pointing out to us the usefulness of the $\mtt$ variable in our context. We also acknowledge L.~Fiorini, S.~Frixione and M.~Volpi for discussions.
This work was supported by the Swiss National Science Foundation under contract No. 200021-116372.

\clearpage





\end{document}